\documentclass[12pt]{article}
\usepackage[dvips]{graphicx}  

\begin{document}

\begin{center}

{\Large \bf A Quantum Cosmology: No Dark Matter,    } 

\vspace{3mm}
 {\Large \bf  Dark Energy nor Accelerating Universe} 

\bigskip

Reginald T. Cahill 

{\small \it  School of Chemistry, Physics and Earth Sciences, Flinders University,
Adelaide 5001, Australia}\\

{\footnotesize  Reg.Cahill@flinders.edu.au}

\begin{abstract}
We show that modelling the universe  as a pre-geometric system with emergent quantum modes, and then constructing the classical limit,  we obtain a new account of  space and gravity that goes beyond Newtonian gravity even in  the non-relativistic limit.  This account does not require dark matter to explain the spiral galaxy rotation curves, and explains as well the observed systematics of black hole masses in spherical star systems, the bore hole $g$ anomalies,  gravitational lensing and so on. As well the dynamics has a Hubble expanding universe solution that gives an excellent parameter-free account of the supernovae and gamma-ray-burst red-shift data, without dark energy or dark matter. The Friedmann-Lema\^{i}tre-Robertson-Walker  (FLRW) metric is derived from this dynamics, but is shown  not satisfy the General Relativity based Friedmann equations.  It is noted that General Relativity  dynamics only permits an expanding flat 3-space solution if the energy density in the pressure-less dust approximation  is non-zero.  As a consequence dark energy and dark matter are required  in this  cosmological model, and as well the prediction of a future exponential  accelerating Hubble expansion. The FLRW  $\Lambda$CDM model  data-based parameter values, $\Omega_\Lambda=0.73$, $\Omega_{DM}=0.27$, are derived within the quantum cosmology model, but are  shown to be merely artifacts of using the  Friedmann equations in fitting the red-shift data.
\end{abstract}
\end{center}

PACS: {98.80.-k, 98.80.Es, 95.35.+d, 95.36.+x

\newpage

\tableofcontents

\section{Introduction}

The current  $\Lambda$CDM standard model of cosmology is based upon General Relativity (GR) as applied to the spatially-flat Friedmann-Lema\^{i}tre-Robertson-Walker  (FLRW) spacetime metric  together with  the Weyl postulate for the energy-momentum density tensor, leading to the Friedmann equations for the 3-space scale factor \cite{Friedmann, Lemaitre, Robertson, Walker}.  Fitting this model to the magnitude-redshift data from supernovae and gamma-ray-burst (GRB) data requires the introduction of dark energy and  dark matter, and a concomitant future exponential acceleration of the universe \cite{PS}.  The dark energy has been most simply interpreted as a cosmological constant $\Lambda$.  Fitting the data gives $\Omega_\Lambda=0.73$, $\Omega_{M}=0.27$, with baryonic matter 
forming only some $\Omega_b=0.05$ of  $\Omega_{M}$, so that the `dark matter' component has $\Omega_{DM}=0.22$.  Hence according to this GR-FLRW\footnote{We use the acronym GR-FLRW when referring to the FLRW metric being determined by GR, giving the Friedmann equations, as the FLRW metric also arises in the context  of a different dynamics discussed herein.} model the universe expansion is determined   mainly by dark energy and cold dark matter, leading to the $\Lambda$CDM label.  A peculiar aspect of the  $\Lambda$CDM model is that the universe can {\it only} expand if the energy density is non-zero, i.e. space itself  cannot  expand without that energy density being present.  This has been a feature of the FLRW dynamics from the beginning of cosmology.   Here we derive a new cosmology which leads to, apart from other numerous tests, an expanding flat 3-space which does not require the presence of energy for that expansion.  This expansion gives a parameter-free fit to the supernovae/GRB data, without invoking dark energy or dark matter.  Nevertheless, if we best-fit the  GR-FLRW  $\Lambda$CDM model to the new cosmology dynamics over the redshift range $z\in\{0,14\}$,  by varying $\Omega_\Lambda$, we obtain   $\Omega_\Lambda=0.73$, $\Omega_{M}=1-\Omega_\Lambda=0.27$. In other words we can predict that fitting the GR-FLRW model to the data will give the parameter values exactly as reported.
However the new cosmology does not predict an accelerating universe; that is merely a spurious consequence of the GR-FLRW model having the wrong functional form for  its Hubble function.   These results change completely our understanding of the evolution of the universe, and of its contents. Basically there is just a very small amount of conventional matter, as indeed deduced from CMB temperature fluctuation data, and a dominant   expanding dynamical 3-space: both of these components of reality may be understood in a unified manner  as aspects of a quantum cosmology model.  

The new cosmology dynamics arises from an information-theoretic modeling of reality which does not assume {\it a priori} any notions about space or quantum matter or even gravity - these phenomena are emergent.  It does however assume that time is modelled as an {\it a priori} stochastic process.  From this pre-geometric system it has been shown that geometry emerges, but  that this is a very complex spatial effect that requires a quantum field theoretic description, namely Quantum Homotopic Field Theory (QHFT).   In this quantum matter corresponds to topological defects of a quantum-foam 3-space.  QHFT is very non-local. We construct here the minimal classical-field description of the dynamical 3-space, and by generalising the Schr\"{o}dinger and  Dirac equations to take account of the dynamical 3-space we find that the quantum matter, in the classical limit, responds to the 3-space dynamics in a manner that we recognise as gravity - that is, we have the first derivation of gravity. As well we find the emergence of the Equivalence Principle.   The minimal model for the 3-space dynamics is found to have two coupling constants, one is the long-known Newtonian $G$, the second is the fine structure constant $\alpha$. This value of $\alpha$ is determined from both bore hole $g$ anomaly data, and from the black hole mass  data.
This suggests the discovery of a  unification in fundamental physics, that the dynamics of space as a quantum-foam  and of quantum electrodynamics are both determined by $\alpha$.

A new dynamics for 3-space must be confirmed by a range of different experiments and observations, and such results are briefly reviewed herein.  We see that the 3-space dynamics gives rise to a description of gravity that  differs from Newtonian gravity - even in the non-relativistic limit.  The differences can be extremely small, as in the solar system, and extremely large as in the case of spiral galaxies and black holes.  The black holes of the new theory have very different  properties from those of GR, and it is this difference that explains the rotation characteristics of spiral galaxies, and so on.  As well a necessary generalisation  of the Maxwell equations leads to a direct and simple account of gravitational light bending and lensing - in the case of light lensing by galactic black holes we note that these black holes generate exceptionally large lensing, which up to now has been explained as being caused by huge quantities of inferred  but undetected `dark matter'.

Because the underlying theory is pre-geometric  non-locality of the dynamics is an expected  emergent property, 
and so the universe is predicted to have a non-local connectivity that exceeds any so far considered. It is suggested  that this is possible explanation for the uniformity of the universe, and so solving the horizon problem.   As well  the 3-space dynamics has a uniformly expanding 3-space, when the EM and baryonic matter energy densities  becomes sufficiently small,  showing that the  universe is flat irrespective of the energy density - this explains the flatness problem. In the GR-FLRW cosmology a flat expanding universe can only arise for an incredibly finely tuned matter  density.  This, it now turns out, was indicating a fundamental flaw in the GR-FLRW  cosmology.

\section{Information-Theoretic Pregeometry}

We consider a bootstrapped self-limited information-theoretic network of `events', labelled primitively by indices $i,j=1,2,3....$, by having a connectivity measure $B_{ij}\in {\cal R}$. The system evolves by  stochastic iterations
\begin{equation}
B_{ij} \rightarrow B_{ij} -a (B + B^{-1})_{ij} + w_{ij},  \mbox{\ \ } i,j=1,2,...,2N;
N \rightarrow \infty
\label{eqn:1}\end{equation}
 where  $w_{ij}$ are
independent random variables for each $ij$ pair and for each iteration and chosen from some probability
distribution. Here $a$ is a parameter the precise value of which should not be critical but which
influences the self-organisational process.   Eqn.(\ref{eqn:1}) behaves as a order-disorder system for emergent patterns.   There is no actual distinction between nodes and links - that is merely an aspect of the bootstrapping system: the emergent system is expected to be fractal, and so nodes then are also to be understood as nothing more  than connection networks.  The term `information' refers to the absence of any notion of substantive matter, and that only {\it patterns} or {\it form}s are actual in this ontology. The matrix inversion $B^{-1}$ involves, in principle, a knowledge of all components of $B$ - amounting to  full self-referencing, while the intrinsic noise $w$ limits the efficacy of that self-referencing.  This limitation to self-referencing has been related to G\"{o}del's theorem re formal logical systems \cite{Book} - essentially we are modelling reality by a system that acknowledges the limits to logic, that a formal syntactical logic system is not suitable for a {\it Theory of Everything}. The iterations amount to a new non-geometric model of time.  Because of the noise $w$ the iterations are irreversible - so modelling a key aspect of the phenomenon of time.  Clearly there is no {\it a priori} notion of geometry, space, quantum matter, etc built into this model.  Nevertheless there is considerable evidence that these phenomena are emergent  \cite{Book}.  The  essential dynamics is that of pattern formation and pattern recognition, and of the preservation over iterations of certain patterns, namely those having a non-trivial fractal  topology.  To briefly give some indication of this self-organisation of persistent patterns we note that at each iteration the $w$ matrix can be considered as having a block-diagonal form by considering only the  rare  large components. Each such block  can be considered to form a random graph structure.  These graphs have an intrinsic approximate geometrical  property \cite{CK} - to characterise that geometry we construct the graph's minimal spanning tree graph. Nagels \cite{Book,Nagels} has shown that the probability of a random connected graph having a spanning tree with  $D_k$ nodes at link count $k$  from some arbitrarily chosen node is given by
\begin{equation}{\cal P}[D,L,N] \propto \frac{p^{D_1}}{D_1!D_2!....D_L!}\prod_{i=1}^{L-1}
(q^{\sum_{j=0}^{i-1}{D_j}})^{D_{i+1}}(1-q^{D_i})^{D_{i+1}}\label{eqn:3}\end{equation}
Here $p=1-q$ is the probability of a link between any two nodes,  $N$ is the number of nodes in the random graph and $L$ is the maximum depth of the tree.   The first indication of  emergent geometry  is obtained by numerically searching for the depth distribution $D_1,D_2,D_3,...$ that maximises ${\cal P}$.  The result of one such computation is shown in figure \ref{fig:gebit}a.  We find that the depth distribution of the minimal spanning tree of the  most probable graph is well fitted by the analytic form
\begin{equation}
D_k\propto \sin^{d-1}\left(\frac{\pi k}{L}\right), \mbox{\ \ \ \  } k=1,2,....,L
\label{eqn:DForm}\end{equation}
\begin{figure}
\hspace{15mm}\parbox{65mm}{\includegraphics[width=64mm,scale=0.15]{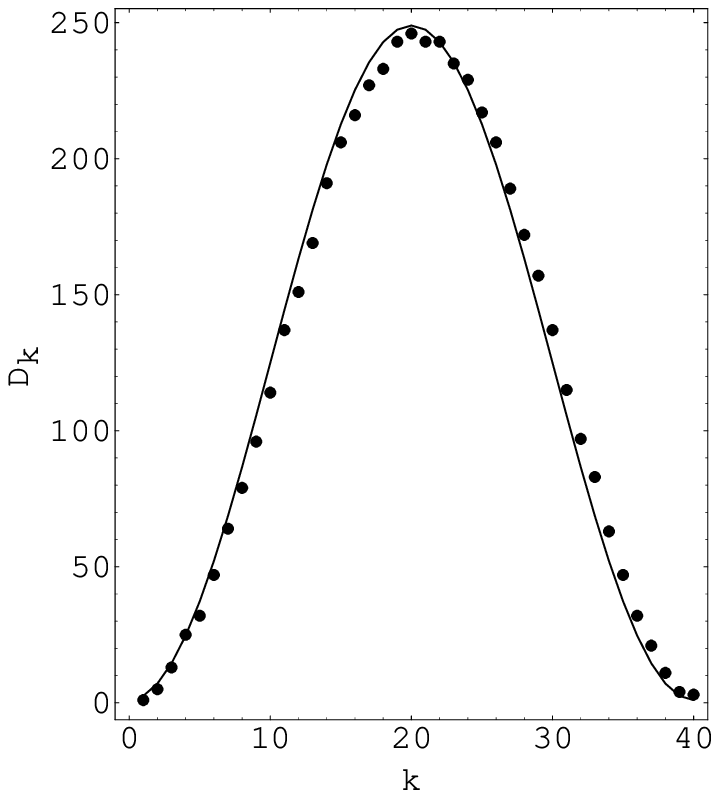}}
\parbox{65mm} {\includegraphics[width=60mm,scale=0.11]{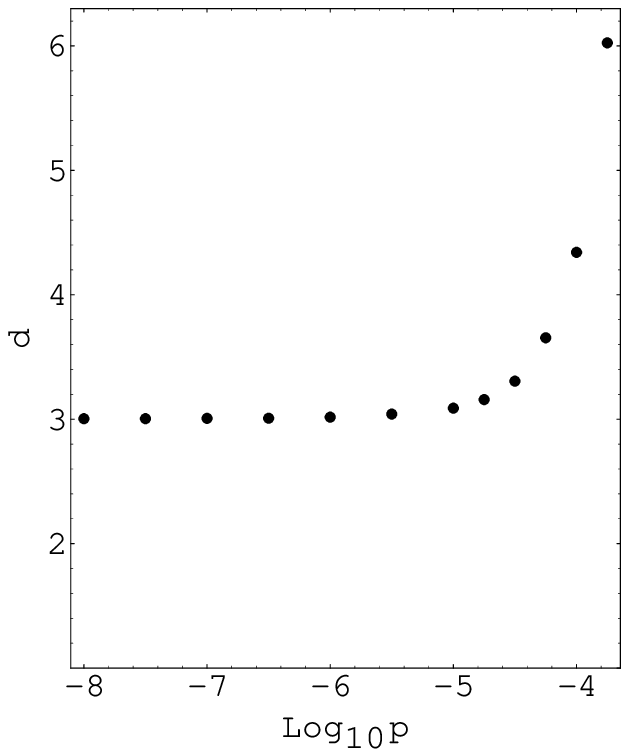}}
\caption{\small 
(a) Points show the $D_k$ set and $L=40$ value found by numerically  maximising ${\cal P}[D,L,N]$
for $\mbox{Log}_{10}p=-6$ for fixed  $N=5000$. Curve shows
$D_k\propto \sin^{d-1}(\frac{\pi k}{L})$ with best fit $d=3.16$ and $L=40$, showing  excellent
agreement, and indicating  embeddability in an $S^3$ with some topological defects. (b) Dimensionality $d$ of
the gebits as a function of  the probability $p$.  The linkage connections have the distribution of a 3D space, but the individual gebit components are closed compact spaces and cannot be embedded in a flat 3D background space. }\label{fig:gebit} 
\end{figure}
The key point is that if the random graph were to be naturally embeddable in a hypersphere $S^d$, then we would expect this form for $D_k$.  As shown in figure \ref{fig:gebit}b this is what emerges: for low values of the probability $p$ we find that $d=3$. This shows that the random matrix that drives (\ref{eqn:1}) is generating primitive geometrical structures  - this is a major insight, and shows that geometry is not necessarily  fundamental to reality, that it is an emergent and approximate aspect.
These `bits of geometry' have been termed {\it gebits}.   The next stage in exposing the dynamics in (\ref{eqn:1}) is that these gebits are polymerised by the inverse $B^{-1}$ process; that is, the gebits are enmeshed or cross-linked.  As this is itself random this next level of structure formation is also subject to a minimal spanning tree analysis, as in (\ref{eqn:DForm}), so the gebits themselves are polymerised into a higher-level hypersphere.  This suggests that (\ref{eqn:1}) generates a fractal geometrical network with the geometrical sub-components linked by discrete homotopies, as suggested by the graphic in figure \ref{fig:Embedd}a. As discussed below it appears that this fractal network of homotopies is describable by a quantum foam.  The gebits are  not preserved - the iterations in (\ref{eqn:1}) essentially bring about their polymerisation and then their decay, so the gebits forming the patterns are continually replaced.  So we have an ongoing processing network  exhibiting a coarse-grained  3-dimensionality.  For the patterns of connected gebits to be preserved, as distinct from individual gebits, the connectivity must posses a non-trivial topology, for then the topology can be preserved  without preserving its individual components. It is suggested that these  preserved structures are the phenomenon known as quantum matter.
This brief account suggests that the information-theoretic network in (\ref{eqn:1}) generates a complex geometric network with preserved  topological defects embedded in an evolving 3-dimensional network which appears to be  what we know of as `space'.  See also the work in \cite{KMS} on a graph theory. In the next section we discuss the observation that if the network  is indeed a quantum foam system formed by quantum homotopies, then we obtain a link to the standard model of particle physics.  The main focus  of this work is, however,  the determination of the  classical-limit dynamics of the emergent quantum foam, for it is different from both the Newtonian and GR treatment of space and gravity.

\section{Quantum Homotopic Field Theory: Emergent Quantum Cosmology}

We suppose here that the lowest level pattern formation arising from (\ref{eqn:1})  is modelled by homotopic mappings between the gebits: this is because the polymerisation of the gebits  is not unique and not formally perfect as in the usual mathematical definition of a homotopy between two compact spaces, and so a functional  over homotopies is perhaps a more appropriate mathematical structure to describe imperfect homotopies.  This suggests that the appropriate mathematical language is that of a Quantum Homotopic Field Theory (QHFT) \cite{Book}.
To construct this QHFT we introduce an appropriate configuration space, namely  all the possible homotopic 
mappings $\pi_{\alpha\beta}: S_\beta \rightarrow S_\alpha$, with 
$S_1,S_2,..$ describing `clean' or topological-defect free gebits -  compact spaces of various types. Then the QHFT
has the form of an iterative functional Schr\"{o}dinger equation for the discrete time-evolution of a wave-functional 
$\Psi[\{\pi_{\alpha\beta}\},t]$
\begin{equation}\Psi[\{\pi_{\alpha\beta}\},t+\Delta
t]= \Psi[\{\pi_{\alpha\beta}\},t]
-iH\Psi[\{\pi_{\alpha\beta}\},t]\Delta t +   {\cal D} [\Psi]\Psi[\{\pi_{\alpha\beta}\},t].
\label{eqn:7}\end{equation}  
This  form arises as it is models the preservation of topologically patterned information, by means of a unitary time
evolution.  The last term ${\cal D}$  is the non-linear minimal manner for retaining randomness in a functional Schr\"{o}dinger equation \cite{Book,QSD}:  unitarity is preserved, in the mean, over  the time evolution, that is, the norm 
\begin{equation}
||\Psi ||=\int\prod_{\alpha\neq\beta}{\cal D}\pi_{\alpha\beta}\mu(\pi_{\alpha\beta})|\Psi[\{\pi_{\alpha\beta}\},t]|^2
\label{eqn:norm}\end{equation}
is invariant under the time evolution  in the mean. Here $\mu(\pi_{\alpha\beta})$ is some suitable integration measure for the homotopy $\pi_{\alpha\beta}$.
 Because of this term  (\ref{eqn:7}) is an irreversible quantum system, and is known to produce wave functional collapse.  Such a term explains the Born measurement metarule in the context of the conventional Schr\"{o}dinger equation \cite{QSD}.    In the above it produces on-going decoherence leading to a quasi-classical 3-space.  The interpretation of the unitary nature of the time evolution is that in general it describes a `bubbling' quantum-foam with the net strength of the connectivities preserved over time, but one without  necessarily forming a quasi-classical 3-space.  Occasionally however  a quasi-classical 3-space arises via essentially a phase-transition - this is  a {\it Big Bang} event.  Once that happens the interaction term ${\cal D}$  continues to maintain that new growing phase. A key result of the work herein  is the determination of the classical-limit dynamics for that process, and its comparison with the experimental and observational data.

 The time step $\Delta t$ in (\ref{eqn:7})  is relative to the scale of the
fractal processes being explicitly described, as we are using a configuration space of mappings between
prescribed gebits.  At smaller scales we would need a smaller value for   $\Delta t$.   
We now consider the form of the hamiltonian $H$.  In \cite{Book}  it was suggested that Manton's
non-linear `elasticity' interpretation \cite{Manton} of the homotopic mappings is appropriate. 
This then suggests that $H$ is the functional operator
\begin{equation}
H=\sum_{\alpha\neq\beta}h\left[\frac{\delta}{\delta
\pi_{\alpha\beta}},\pi_{\alpha\beta}\right],
\label{eqn:8}\end{equation}
where $h[\frac{\delta}{\delta \pi},\pi]$ is the (quantum) Skyrme Hamiltonian functional operator for the system based
on  making fuzzy the  mappings
$\pi: S \rightarrow \Sigma$, by having $h$ act on wave-functionals of the form $\Psi[\pi(x);t]$. Then $H$
is the sum of pairwise  embedding or homotopy hamiltonians. The corresponding functional Schr\"{o}dinger
equation \index{functional Schr\"{o}dinger equation} would simply  describe the time evolution of quantised
Skyrmions.  We shall not give the explicit form of $h$ as it is complicated, but wait to present
the associated induced action. 

In the absence of the non-linear ${\cal D}$ terms the time evolution in (\ref{eqn:7}) can be formally written as a
functional integral
\begin{equation}
\Psi[\{\pi\};t']=\int\prod_{\alpha\neq\beta}{\cal
D}\tilde{\pi}_{\alpha\beta}e^{iS[\{\tilde{\pi}\}]}
\Psi[\{\pi\};t],
\label{eqn:9}\end{equation}
where, using the continuum $t$ limit notation, the action is a sum of pairwise actions,
\begin{equation}
S[\{\tilde{\pi}\}]=\sum_{\alpha\neq\beta}S_{\alpha\beta}[\tilde{\pi}_{\alpha\beta}],
\label{eqn:10}\end{equation}
\begin{eqnarray}
S_{\alpha\beta}[\tilde{\pi}]&=&\int_t^{t'}dt''\int d^nx\sqrt{ -g} \left[ \frac{1}{2}\mbox{Tr}(\partial_\mu
\tilde{U}\tilde{U}^{-1}\partial^\mu
\tilde{U}\tilde{U}^{-1})\right. + \nonumber \\
&&\mbox{\ \ \ \ \ \ \ \ \ \ \ \ \ \ \ \ \ \ \ \ \ \ \ \ \  }\left. \frac{1}{16} \mbox{Tr}[\partial_\mu \tilde{U}\tilde{U}^{-1},\partial^\nu
\tilde{U}\tilde{U}^{-1}]^2\right],
\label{eqn:11}\end{eqnarray}
and the now time-dependent (indicated by the tilde symbol) mappings $\tilde{\pi}$ are parametrised by
$\tilde{U}(x,t)$, $\tilde{U}\in S_\alpha$. The metric $g_{\mu\nu}$ is that of the $n$-dimensional base space,
$S_\beta$, in
$\pi_{\alpha,\beta}: S_\beta
\rightarrow  S_\alpha$. As usual in the functional integral formalism the functional derivatives in the quantum
hamiltonian, in (\ref{eqn:8}), now manifest as the time components $\partial_0$ in (\ref{eqn:11}),  so now
(\ref{eqn:11}) has the form of a `classical' action, and we see the emergence of `classical' fields\index{classical
fields}, though the emergence of `classical' behaviour\index{classical behaviour} is a more complex  process and requires the retention of the ${\cal D}$ term in (\ref{eqn:7}). 

A key observation is that the quantum system (\ref{eqn:9})-(\ref{eqn:11})  is essentially that of the Nambu-Goldstone modes  of the standard model of particle physics - by a functional change of field variables these modes may be related to a QFT involving fermionic and vector gauge fields, which are the conjectured preonic fermions and hypercolour vector fields. \cite{Book}.  It is not claimed that this is a formal derivation of the standard model of particle physics from (\ref{eqn:1}) - that will require much further study. See also the related work in \cite{BT}.  Nevertheless it is suggestive of how a self-organising  and self-limited
information-theoretic system may lead to emergent  behaviour describable by a quantum-theoretic formalism, and achieve that with a very generic bootstrapping  system as in  (\ref{eqn:1}), which does not assume any emergent pattern dynamics, such as space or quantum matter,  yet with  the emergence of such phenomena, and in a unified manner as  (\ref{eqn:9}) describes both a quantum-foam dynamical 3-space as well as quantum matter.  If, as suggested, a phase transition to a growing dynamical 3-space system  arises, then self-organised criticality would arise, and then the details of the generic system in (\ref{eqn:1}) would be hidden. In the next section we construct a minimal phenomenological classical  dynamics for this quantum-foam space and discover that it predicts the phenomenon of gravity, but  possessing dynamical aspects that account for the effects that, until now, have required the introduction of `dark matter' and `dark energy'.  

\begin{figure}[t]
\hspace{20mm}\parbox{60mm}{\hspace{10mm}\includegraphics[width=55mm]{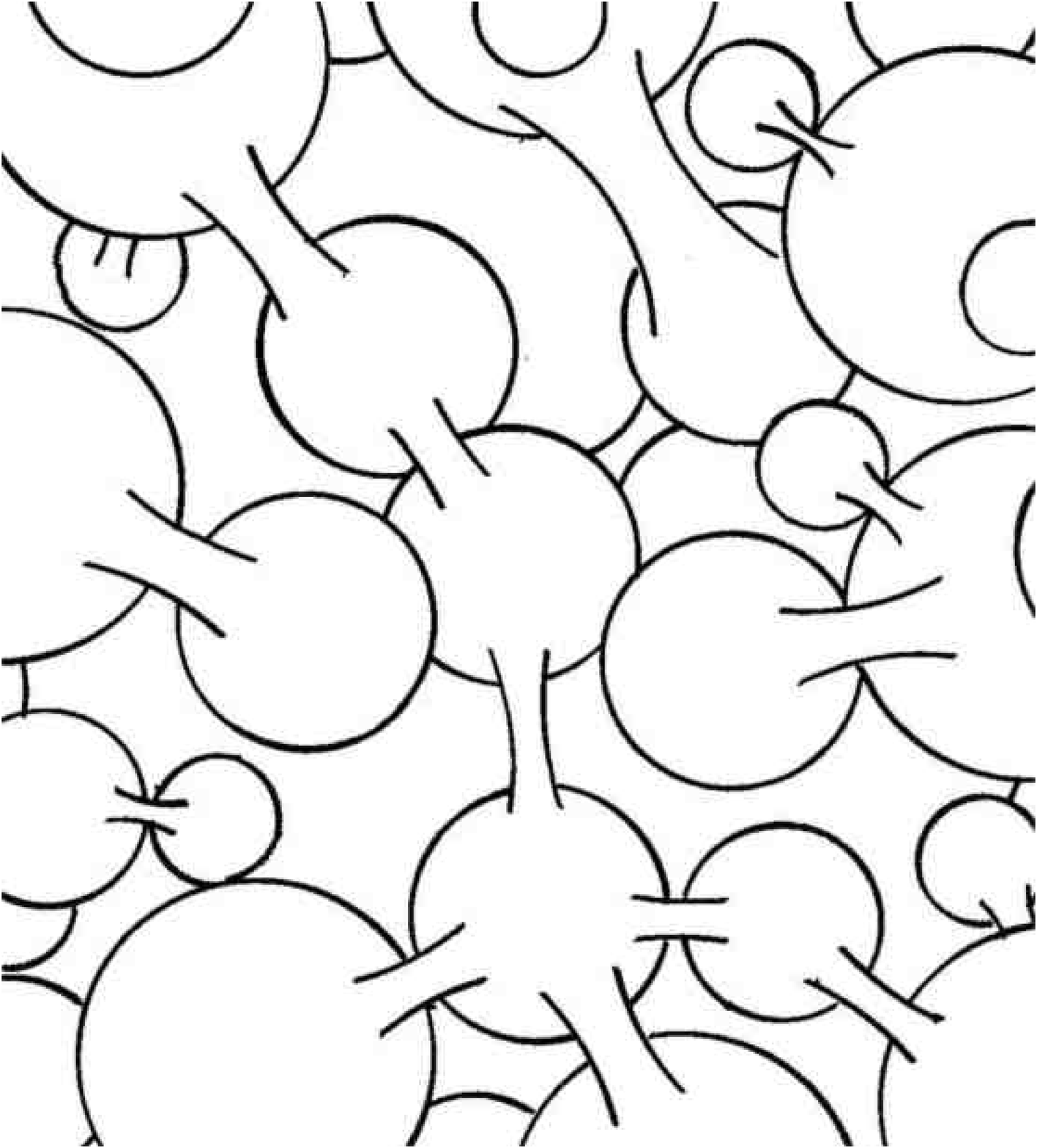}}\,
\parbox{70mm}{\vspace{0mm}\,\parbox{60mm}{\hspace{10mm}\includegraphics[width=60mm]{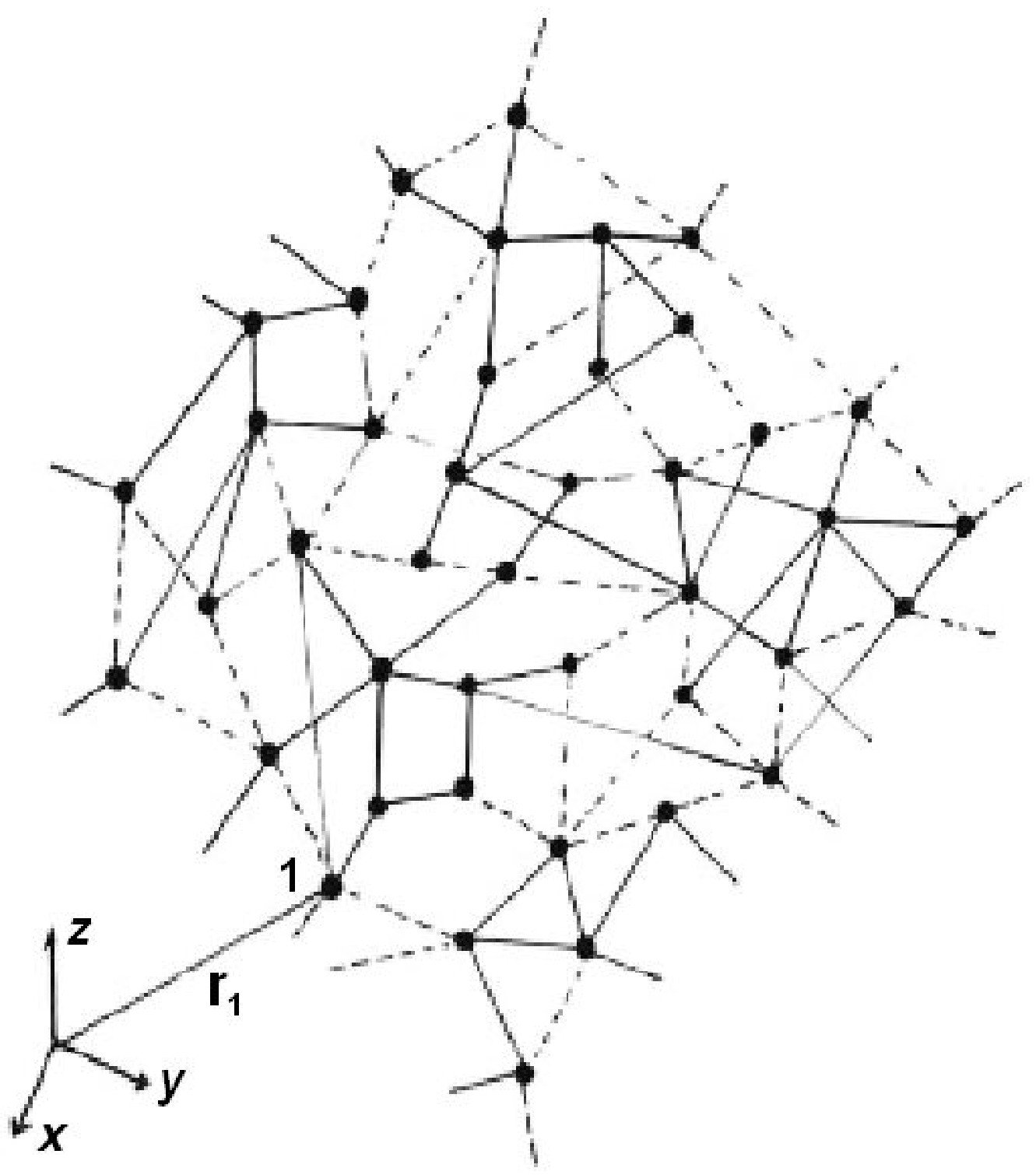}}\,}
\caption{\footnotesize{  This is an iconic representation of how the iterative scheme in (\ref{eqn:1}) generates a  homotopic network (left) describable as a quantum foam. Its skeletal structure  has its inherent approximate 3-dimensional connectivity displayed by an embedding in a mathematical   space, such as an $E^3$ or an $S^3$, as shown on the right.  The embedding space is not real - it is purely a mathematical artifact. Nevertheless this embeddability helps determine the minimal dynamics for the network, as in (\ref{eqn:E1}).   The dynamical space is not an ether model, as the embedding space does not exist.}}
\label{fig:Embedd}
\end{figure}

\section{From Quantum Cosmology to Classical Dynamical 3-Space}

The information-theoretic approach to modelling reality  leads to an emergent structured quantum-foam `space'  which is 3-dimensional and dynamic, but where the 3-dimensionality is only approximate, in that if we ignore non-trivial topological aspects of the quantum foam, then it may be embedded in a 3-dimensional  geometrical manifold \cite{CPP}.  Here the space is a real existent discrete but fractal network of relationships or connectivities,  but the embedding space is purely a mathematical way of characterising the 3-dimensionality of the network.  This is illustrated by the  skeletal representation of the quantum foam in figure \ref{fig:Embedd}b - this is not necessarily local in that significant linkages can manifest between distant regions.  Embedding the network in the embedding space is very arbitrary; we could equally well rotate the embedding or use an embedding that has the network translated or translating.  These general requirements  then dictate the minimal dynamics for the actual network, at a phenomenological level.  To see this we assume  at a coarse grained level that the dynamical patterns within the network may be described by a velocity field ${\bf v}({\bf r},t)$, where ${\bf r}$ is the location of a small region in the network according to some arbitrary embedding.  The 3-space velocity field has been observed in at least 9 experiments  
\cite{Book}. 
For simplicity we assume here that the global topology of the network   is not significant for the local dynamics, and so we embed in an $E^3$, although a generalisation to an embedding in $S^3$ is straightforward and might be relevant to cosmology.  The minimal dynamics is then obtained by  writing down the sum of the only three lowest-order zero-rank tensors, of dimension $1/T^2 $, that are invariant under translation and rotation, giving
\begin{equation}
\nabla.\left(\frac{\partial {\bf v} }{\partial t}+({\bf v}.{\bf \nabla}){\bf v}\right)
+\frac{\alpha}{8}(tr D)^2 +\frac{\beta}{8}tr(D^2)=
-4\pi G\rho
\label{eqn:E1}\end{equation}
\begin{equation}D_{ij}=\frac{1}{2}\left(\frac{\partial v_i}{\partial x_j}+
\frac{\partial v_j}{\partial x_i}\right)
\label{eqn:E1b}\end{equation}
where $\rho({\bf r},t)$ is an effective matter density that may  correspond to various energy densities. The embedding space coordinates provide a coordinate system or frame of reference that is convenient to describing the velocity field, but which is not real.  

We see that there are only four possible terms, and so we need at most three possible constants to describe the dynamics of space: $G, \alpha$ and $\beta$. $G$ turns out  to be Newton's gravitational constant, and describes the rate of non-conservative flow of space into matter.  To determine the values of $\alpha$ and $\beta$ we must, at this stage, turn to experimental and observational data.  
However most data involving the dynamics of space is obtained by detecting the so-called gravitational  acceleration of matter, although increasingly light bending is giving new information.  Now the acceleration ${\bf a}$ of the dynamical patterns in space is given by the Euler or convective expression
\begin{equation}
{\bf a}({\bf r},t)= \lim_{\Delta t \rightarrow 0}\frac{{\bf v}({\bf r}+{\bf v}({\bf r},t)\Delta t,t+\Delta
t)-{\bf v}({\bf r},t)}{\Delta t} 
=\frac{\partial {\bf v}}{\partial t}+({\bf v}.\nabla ){\bf v}
\label{eqn:E3}\end{equation} 
and this appears in one of the terms in (\ref{eqn:E1}). As shown  in \cite{Schrod} and discussed later in Sect. \ref{sect:acceln} the acceleration  ${\bf g}$ of quantum matter is identical to this acceleration, apart from vorticity and relativistic effects, and so the gravitational acceleration of matter is also given by (\ref{eqn:E3}).

Outside of a spherically symmetric distribution of matter,  of total mass $M$, we find that one solution of (\ref{eqn:E1}) is the velocity in-flow field  given by
\begin{equation}
{\bf v}({\bf r})=-\hat{{\bf r}}\sqrt{\frac{2GM(1+\frac{\alpha}{2}+..)}{r}}
\label{eqn:E4}\end{equation}
but only when $\beta=-\alpha$,  for only then is the acceleration of matter, from (\ref{eqn:E3}), induced by this in-flow of the form
\begin{equation}
{\bf g}({\bf r})=-\hat{{\bf r}}\frac{GM(1+\frac{\alpha}{2}+..)}{r^2}
\label{eqn:E5}\end{equation}
 which  is Newton's Inverse Square Law of 1687 \cite{Newton}, but with an effective  mass $M(1+\frac{\alpha}{2}+..)$ that is different from the actual mass $M$.  So the success of Newton's law in the solar system, based on Kepler's analysis, informs us that  $\beta=-\alpha$ in (\ref{eqn:E1}). But we also see modifications coming from the 
$\alpha$-dependent terms.

In general because (\ref{eqn:E1}) is a scalar equation it is only applicable for vorticity-free flows $\nabla\times{\bf v}={\bf 0}$, for then we can write ${\bf v}=\nabla u$, and then (\ref{eqn:E1}) can always be solved to determine the time evolution of  $u({\bf r},t)$ given an initial form at some time  $t_0$.
The $\alpha$-dependent term in (\ref{eqn:E1})  (with now $\beta=-\alpha$) and the matter acceleration effect, now also given by (\ref{eqn:E3}),   permits   (\ref{eqn:E1})   to be written in the form
\begin{equation}
\nabla.{\bf g}=-4\pi G\rho-4\pi G \rho_{DM},
\label{eqn:E7}\end{equation}
where
\begin{equation}
\rho_{DM}({\bf r},t)\equiv\frac{\alpha}{32\pi G}( (tr D)^2-tr(D^2)),  
\label{eqn:E7b}\end{equation}
which  is an effective matter density, not necessarily non-negative,  that would be required to mimic the
 $\alpha$-dependent spatial self-interaction dynamics. The Newtonian coupling constant $G$ is included in the definition of $\rho_{DM}$ only so that its role as an effective matter density can be illustrated - the  $\alpha$ dynamics does not involves $G$.  
 Then (\ref{eqn:E7}) is the differential form for Newton's law of gravity but with an additional non-matter effective matter density.  So we label this as $\rho_{DM}$ even though no matter is involved \cite{alpha,DM}, as this effect has been shown to explain the so-called `dark matter' effect in spiral galaxies, bore hole $g$ anomalies, and the systematics of galactic black hole masses.  
 
 The spatial dynamics  is non-local.  Historically this was first noticed by Newton who called it action-at-a-distance. To see this we can write  (\ref{eqn:E1}) as an integro-differential equation
 \begin{equation}
 \frac{\partial {\bf v}}{\partial t}=-\nabla\left(\frac{{\bf v}^2}{2}\right)+G\!\!\int d^3r^\prime
 \frac{\rho_{DM}({\bf r}^\prime, t)+\rho({\bf r}^\prime, t)}{|{\bf r}-{\bf r^\prime}|^3}({\bf r}-{\bf r^\prime})
 \label{eqn:E8}\end{equation}
This shows a high degree of non-locality and non-linearity, and in particular that the behaviour of both $\rho_{DM}$ and $\rho$ manifest at a distance irrespective of the dynamics of the intervening space. This non-local behaviour is analogous to that in quantum systems and may offer a resolution to the horizon problem. As well the dynamics involving  $\rho_{DM}$ manifests at a a distance  to a scale independent of $G$, because of the $1/G$ coefficient in $\rho_{DM}$, as noted above, and so `gravitational wave' effects caused by distant activity are predicted to be much large than predicted by GR.

\begin{figure}
\hspace{35mm}\includegraphics[scale=0.3]{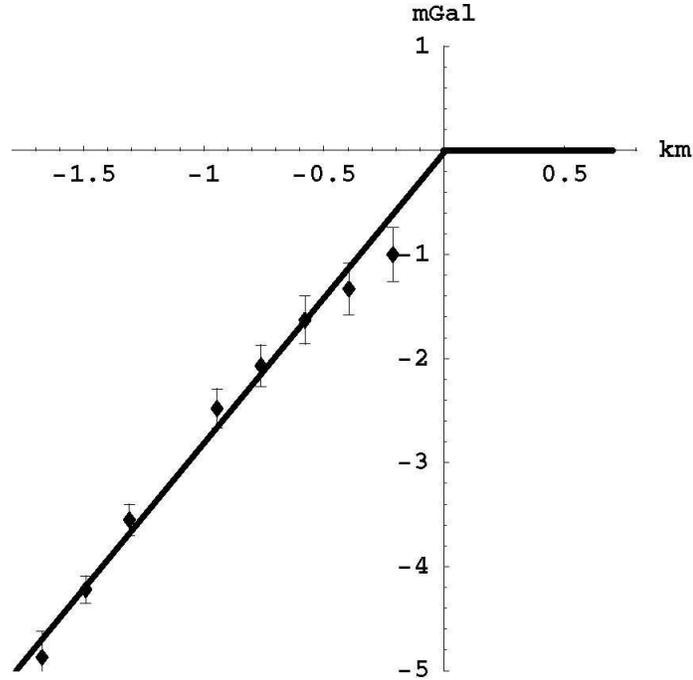}
{\caption{\small{  The data shows the gravity residuals for the Greenland Ice Shelf \cite{Ander89} Airy measurements of
the $g(r)$  profile,  defined as $\Delta g(r) = g_{Newton}-g_{observed}$, and measured in mGal (1mGal $ =10^{-3}$ cm/s$^2$)
and   plotted against depth in km. The borehole effect is that Newtonian
gravity and the new theory differ only beneath the surface, provided that the measured above-surface gravity gradient 
is used in  both theories.  This then gives the horizontal line above the surface. Using (\ref{eqn:E6}) we obtain
$\alpha^{-1}=137.9 \pm  5$ from fitting the slope of the data, as shown. The non-linearity  in the data arises from
modelling corrections for the gravity effects of the   irregular sub ice-shelf rock  topography. The ice density is 920 kg/m$^3$.}}
\label{fig:Greenland}}
\end{figure}

\begin{figure}[t]
\hspace{10mm}\,\hspace{10mm}\parbox{70mm}{\includegraphics[width=60mm,scale=0.2]{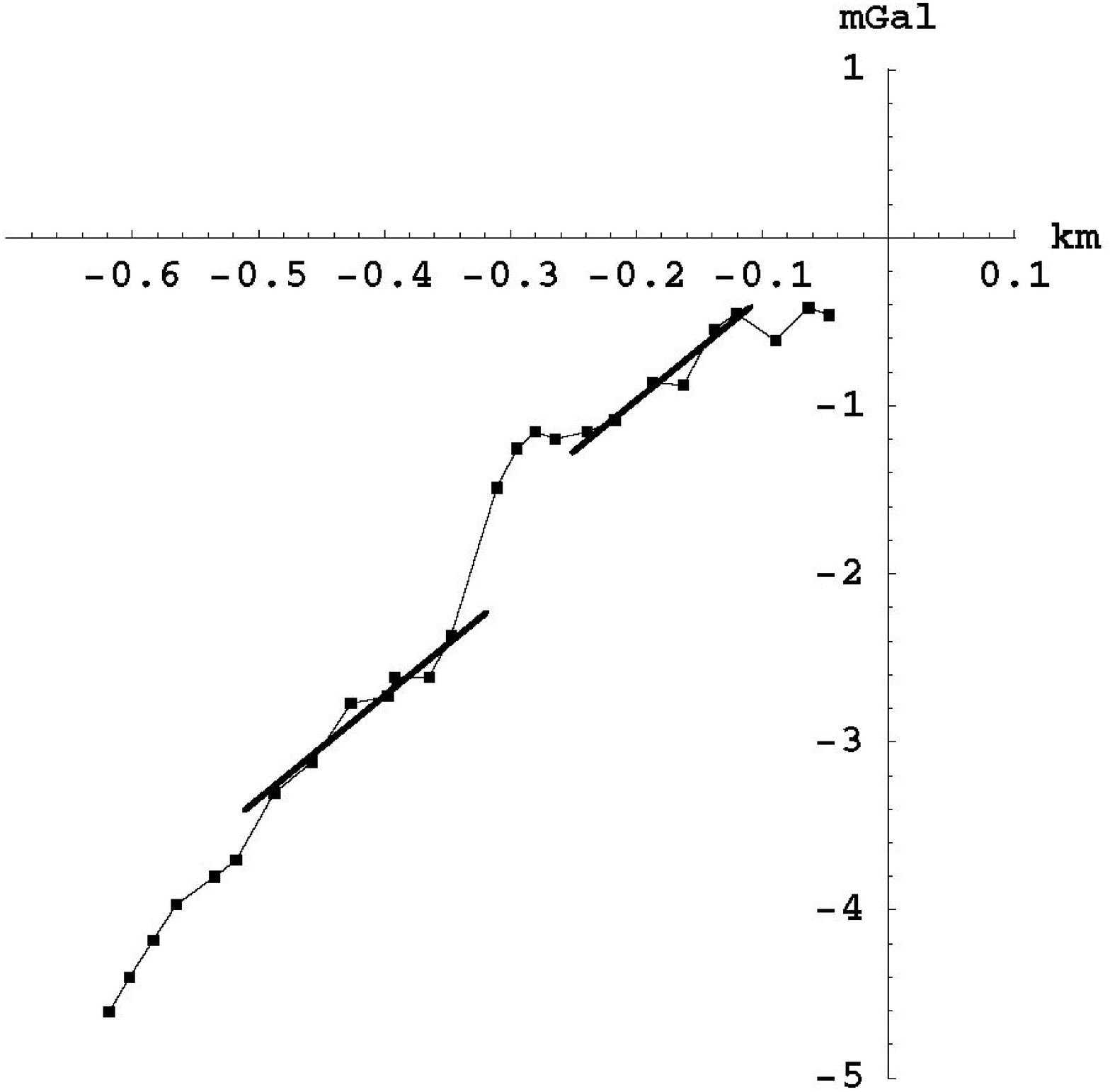}}
\parbox{70mm} {\includegraphics[width=60mm,scale=0.2]{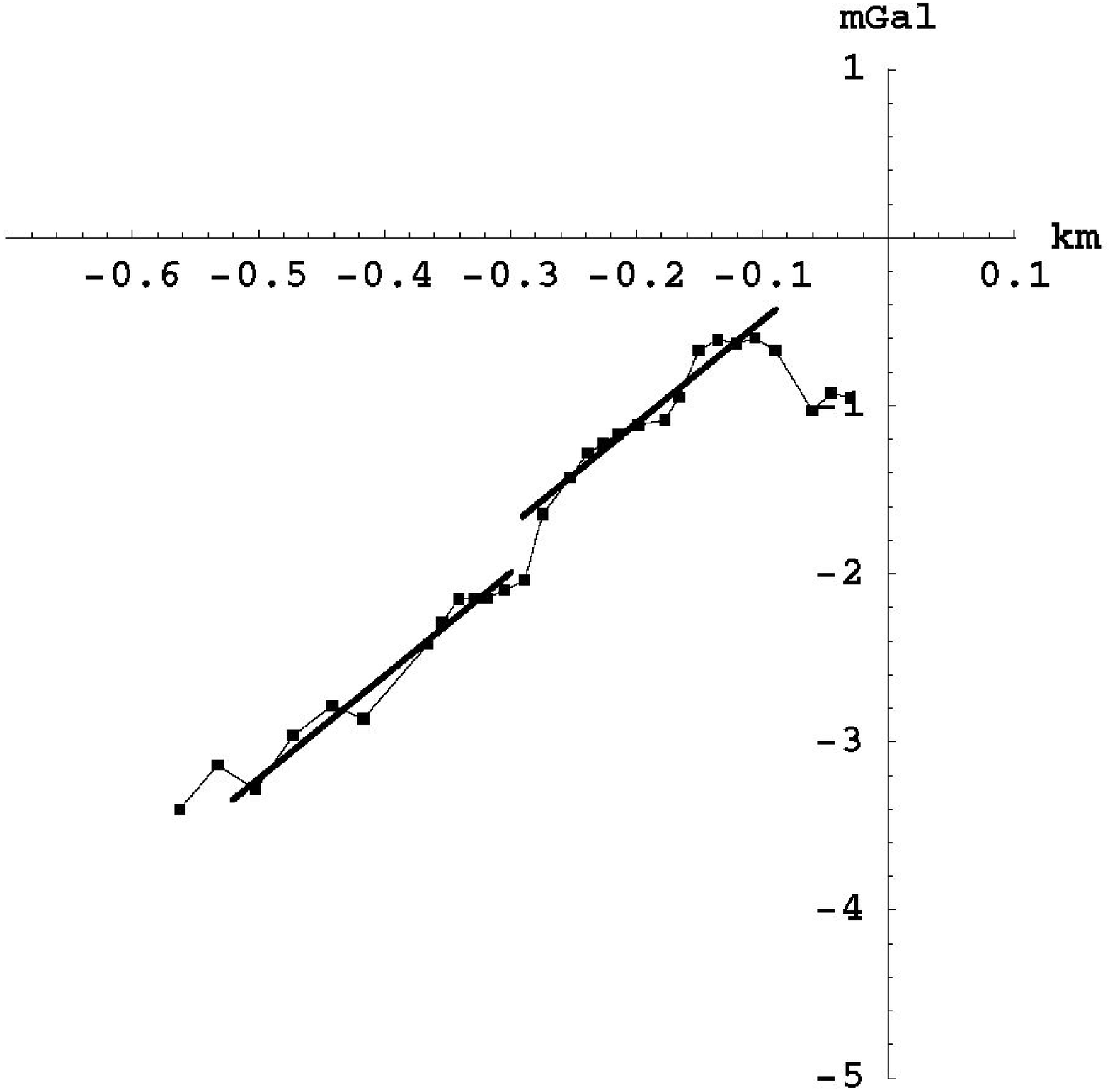}}
{\caption{\small{  Gravity residuals  $\Delta g(r) $ from two of the Nevada bore hole experiments \cite{Thomas90} that give a best fit of $\alpha^{-1}=136.8\pm 3$ on using (\ref{eqn:E6}). Some layering of the rock is evident. The rock density is 2000 kg/m$^3$ in the linear regions.}}
\label{fig:Nevada}}
\end{figure}
 
 \section{Bore Hole Anomaly: Fine Structure Constant}
 
 A recent discovery \cite{alpha, DM} has been that experimental data from the bore hole $g$ anomaly has revealed that $\alpha$ is the fine structure constant, to within experimental errors: $\alpha=e^2/\hbar c \approx 1/137.04$. This observed anomaly is that $g(r)$ does not decrease as rapidly as predicted by Newtonian gravity or GR as we descend down a bore hole.  
 Consider the case where we have a spherically symmetric matter distribution, at rest on average with respect to distant space, and that the in-flow is time-independent and radially symmetric.  Then (\ref{eqn:E1})  can now be written in the form,  with $v^\prime=dv(r)/dr$, 
 \begin{equation}
 vv^{\prime\prime}+2\frac{vv^\prime}{r} +(v^\prime)^2  =-4\pi G\rho(r)-4\pi G \rho_{DM}(v(r)), 
\label{eqn:Eradial}
\end{equation} 
where now
 \begin{equation}
\rho_{DM}(r)= \frac{\alpha}{8\pi G}\left(\frac{v^2}{2r^2}+ \frac{vv^\prime}{r}\right).
\label{eqn:E10}\end{equation}
The dynamics in (\ref{eqn:Eradial}) and (\ref{eqn:E10}) gives
 the anomaly to be
 \begin{equation}
 \Delta g=2\pi\alpha G \rho d +O(\alpha^2)
 \label{eqn:E6}\end{equation}
where $d$ is the depth and $\rho$ is the density, being that of glacial ice in the case of the Greenland Ice Shelf experiments \cite{Ander89}, or that of rock in the Nevada test site experiment \cite{Thomas90}. Clearly (\ref{eqn:E6}) permits the value of $\alpha$ to be determined from the data, giving  $\alpha=1/ (137.9 \pm 5)$ from the Greenland data, and  $\alpha=1/(136.8\pm 3)$ from the Nevada data; see Figs. \ref{fig:Greenland} and \ref{fig:Nevada}.  Note that the density  $\rho$ in (\ref{eqn:E6}) is very different for these two experiments, showing that the extracted value $\alpha$  $\approx 1/137$ is robust.

\section{Minimal and Non-Minimal Black Holes: Fine Structure Constant}

\begin{figure}[t]
\hspace{15mm}\includegraphics[scale=0.4]{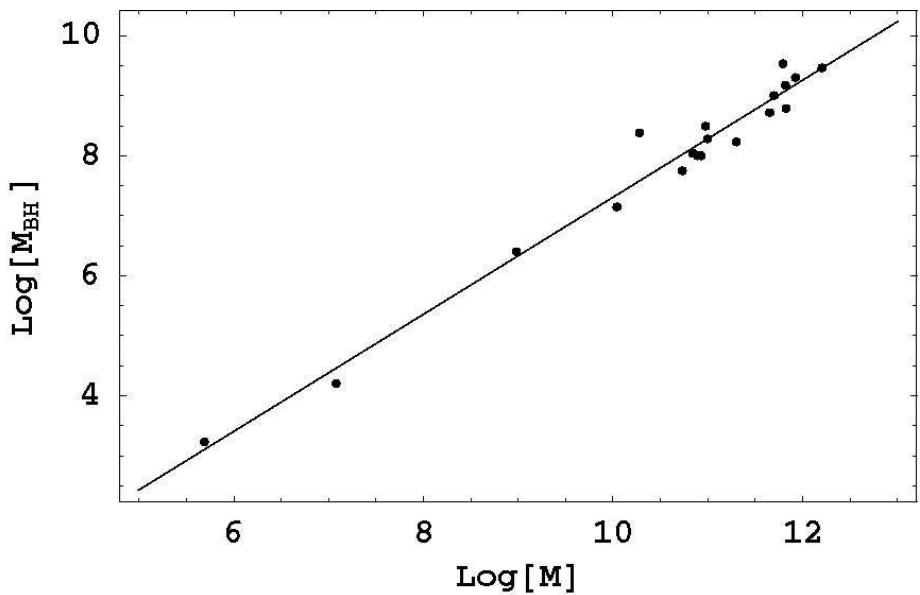}
\vspace{-4mm}
\caption{\small{The data shows $\mbox{Log}_{10}[M_{BH}]$ for the black hole masses $M_{BH}$  for
a variety of spherical matter systems, from Milky Way galactic clusters to spherical galaxies, with masses $M$, plotted against 
$\mbox{Log}_{10}[M]$, in solar masses $M_0$.  The straight line is the prediction from (\ref{eqn:bhmasses}) with $\alpha=1/137$. See \cite{newBH} for references to the data. 
  \label{fig:blackholes}}}
\end{figure}

Eqn.(\ref{eqn:Eradial}) with $\rho=0$ has  exact analytic `black hole' solutions, given by (\ref{eqn:vexactb}) without the $1/r$  term.  There are two classes of black hole solutions - they are distinguished by how they relate to the surrounding matter. The class of minimal black holes is completely induced by the  surrounding distribution of matter.   For a spherically symmetric distribution of matter we find by iterating (\ref{eqn:Eradial}) and then from  (\ref{eqn:E10}) that the total effective black hole  mass is
\begin{equation}
M_{BH}=M_{DM} = 4\pi\int_0^\infty r^2\rho_{DM}(r)dr = \frac{\alpha}{2}M+O(\alpha^2)
\label{eqn:bhmasses}\end{equation}
This solution is applicable to the black holes at the centre of spherical star systems, where we identify $M_{DM}$ as $M_{BH}$.   For these black holes the acceleration $g$ outside of the matter decreases as $1/r^2$. So far black holes in 19  spherical star systems have been detected and together their masses are plotted in 
figure \ref{fig:blackholes} and compared with (\ref{eqn:bhmasses}), giving again $\alpha=1/137$ \cite{galaxies,newBH}.   These solutions are called `black holes' because they posses an event horizon that forbids the escape of EM radiation and matter, but that they are very different from the putative `black holes' of GR. Clearly GR cannot predict the mass relation in (\ref{eqn:bhmasses}) as the GR dynamics does not involve $\alpha$.  The second class of black hole solutions is called non-minimal.  These come into existence before subsequently attracting matter.  These  black holes may be primordial in that they formed directly as a consequence of the big bang before stars and galaxies, and indeed may have played a critical role in the precocious formation of  galaxies.  These black holes are responsible for both the rapid in-fall of matter to form rotating spiral galaxies, and also for  non-Keplerian rotation characteristics of these galaxies, as discussed next.  It is significant that the bore hole, black hole and (next) the spiral galaxy rotation effects are all caused by the non-local dynamics from the $\alpha$-dynamics - and so are indicative of the non-local quantum effects of the quantum cosmology.

\section{Spiral Galaxy Rotation Anomaly: Fine Structure Constant}

\begin{figure}[t]
\hspace{30mm}\includegraphics[scale=1.2]{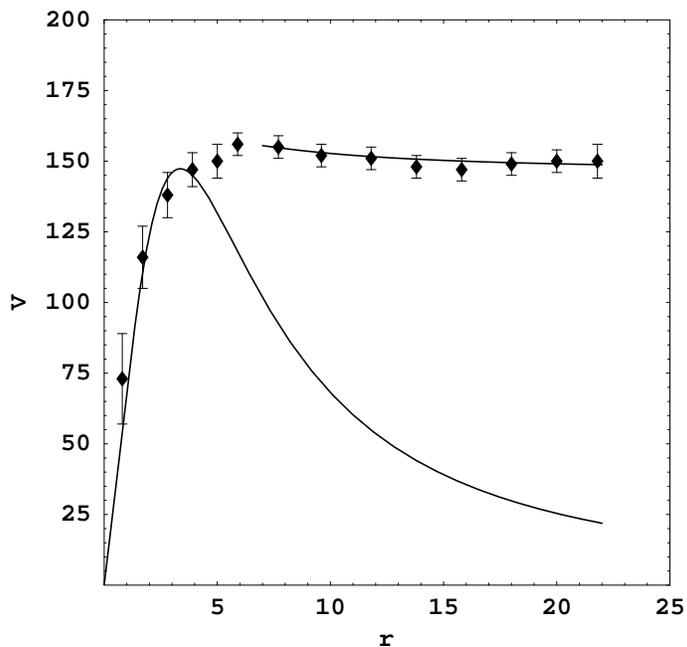}
\vspace{-4mm}
\caption{\small {Data shows the non-Keplerian rotation-speed curve $v_O$ for the spiral galaxy NGC 3198 in km/s plotted
against radius in kpc/h. Lower curve is the rotation curve from the Newtonian theory  for an
exponential disk, which decreases asymptotically like $1/\sqrt{r}$. The upper curve shows the asymptotic form from
(\ref{eqn:vorbital}), with the decrease $\sim 1/r$ determined by the small value of $\alpha$.  This asymptotic form is caused by
the primordial black holes at the centres of spiral galaxies, and which play a critical role in their formation. The
spiral structure is caused by the rapid in-fall towards these primordial black holes.}
\label{fig:NGC3198}}\end{figure}

The black hole solutions of   (\ref{eqn:Eradial}) give a direct explanation for the spiral galaxy rotation anomaly.    For a non-spherical system numerical solutions of (\ref{eqn:E1}) are required, but sufficiently far from the centre we find an exact non-perturbative two-parameter class of analytic solutions
\begin{equation}
v(r) = K\left(\frac{1}{r}+\frac{1}{R_s}\left(\frac{R_s}{r}  \right)^{\displaystyle{\frac{\alpha}{2}}}  \right)^{1/2}
\label{eqn:vexactb}\end{equation}
where $K$ and $R_s$ are arbitrary constants in the $\rho=0$ region, but whose values are determined by matching to the solution in the matter region. Here $R_s$ characterises the length scale of the non-perturbative part of this expression,  and $K$ depends on $\alpha$, $G$ and details of the matter distribution.   From (\ref{eqn:E5})  and (\ref{eqn:vexactb}) we obtain a replacement for  the Newtonian  `inverse square law' ,
\begin{equation}
g(r)=\frac{K^2}{2} \left( \frac{1}{r^2}+\frac{\alpha}{2rR_s}\left(\frac{R_s}{r}\right)
^{\displaystyle{\frac{\alpha}{2}}} 
\right),
\label{eqn:gNewl}\end{equation}
in the asymptotic limit.     The non-Newtonian part of this acceleration is caused by presence of a primordial `black hole' at the centre of the galaxy, about which the galaxy formed: in general the `black holes' from   (\ref{eqn:Eradial}) have an acceleration $g  \sim 1/r$, and very unlike the form $g \sim 1/r^2$ for the putative black holes of GR.  The centripetal acceleration  relation for circular orbits 
$v_O(r)=\sqrt{rg(r)}$  gives  a `universal rotation-speed curve'
\begin{equation}
v_O(r)=\frac{K}{2} \left( \frac{1}{r}+\frac{\alpha}{2R_s}\left(\frac{R_s}{r}\right)
^{\displaystyle{\frac{\alpha}{2}}} 
\right)^{1/2}
\label{eqn:vorbital}\end{equation}
 Because of the $\alpha$ dependent part this rotation-velocity curve  falls off extremely slowly with $r$, as is indeed observed for spiral galaxies. An example is shown in figure \ref{fig:NGC3198}. It was the inability of the  Newtonian  and Einsteinian gravity theories to explain these observations that led to the  notion of `dark matter'.   Note that in the absence of the $\alpha$-dynamics, the rotation-speed curve reduces to the Keplerian form.  Nevertheless  it is not clear if the  form in (\ref {eqn:vorbital}) could be used to determine the value of $\alpha$ from the extensive data set of spiral galaxy rotation curves because of observational errors and intrinsic non-systematic variations  in individual galaxies, unlike the data from bore holes and black holes which give independent but consistent determinations for the value of $\alpha$. We see that the 3-space  dynamics  (\ref{eqn:E1}) gives a unified account of both the `dark matter' problem and the properties of `black holes'.

\section{Generalised  Maxwell Equations: Gravitational Lensing}

 We must   generalise the Maxwell equations so that the electric and magnetic  fields are excitations within the dynamical 3-space, and not of the embedding space.  The minimal form in the absence of charges and currents is 
 \begin{eqnarray}
\displaystyle{ \nabla \times {\bf E}}&=&\displaystyle{-\mu\left(\frac{\partial {\bf H}}{\partial t}+{\bf v.\nabla H}\right)},
 \mbox{\ \ \ }\displaystyle{\nabla.{\bf E}={\bf 0}},  \nonumber \\
 \displaystyle{ \nabla \times {\bf H}}&=&\displaystyle{\epsilon\left(\frac{\partial {\bf E}}{\partial t}+{\bf v.\nabla E}\right)},
\mbox{\ \ \  }\displaystyle{\nabla.{\bf H}={\bf 0}}\label{eqn:E18}\end{eqnarray}
which was first suggested by Hertz in 1890  \cite{Hertz}, but with ${\bf v}$ then being only a constant vector field. As easily determined  the speed of EM radiation is now $c=1/\sqrt{\mu\epsilon}$ with respect to the 3-space.
To see this we  find plane wave solutions for (\ref{eqn:E18}):
\begin{equation}
{\bf E}({\bf r},t)={\bf E}_0e^{i({\bf k}.{\bf r}-\omega t)} \mbox{\ \ \ \  } {\bf H}({\bf r},t)={\bf H}_0e^{i({\bf k}.{\bf r}-\omega t)}
\label{eqn:pw}\end{equation}
with
\begin{equation}
\omega({\bf k},{\bf v})=c|\vec{{\bf k}}| +{\bf v}.{\bf k} \mbox{ \ \ \  where \ \ \  } c=1/\sqrt{\mu\epsilon}
\label{eqn:omega}\end{equation}
Then the EM group velocity is
\begin{equation}
{\bf v}_{EM}=\vec{\nabla}_k\omega({\bf k},{\bf v})=c\hat{\bf k}+{\bf v}
\label{eqn:groupv}\end{equation}
So the velocity of EM radiation ${\bf v}_{EM}$ has magnitude  $c$ only with respect to the space, and in general not with respect to the observer if the observer is moving through space.  

The time-dependent and inhomogeneous  velocity field causes the refraction of EM radiation. This can be computed by using the Fermat least-time approximation. Then the EM ray paths  ${\bf r}(t)$ are determined by minimising  the elapsed travel time:
\begin{equation}
\tau=\int_{s_i}^{s_f}\frac{ds\displaystyle{|\frac{d{\bf r}}{ds}|}}{|c\hat{{\bf v}}_R(s)+{\bf v}(\bf{r}(s),t(s)|}
\mbox{ \ \ with \ \ }
{\bf v}_R=\left(  \frac{d{\bf r}}{dt}-{\bf v}(\bf{r}(t),t)\right)
\label{eqn:lighttime}\end{equation}
by varying both ${\bf r}(s)$ and $t(s)$, finally giving ${\bf r}(t)$. Here $s$ is a path parameter, and ${\bf v}_R$ is a 3-space tangent vector for the path.

In particular the in-flow in (\ref{eqn:E4}) causes a refraction effect of light passing close to the sun, with the angle of deflection given by
\begin{equation}
\delta=2\frac{v^2}{c^2}=\frac{4GM(1+\frac{\alpha}{2}+..)}{c^2d}
\label{eqn:E19}\end{equation}
where $v$ is the in-flow speed at distance $d$  and $d$ is the impact parameter, here the radius of the sun. Hence the  observed deflection of $8.4\times10^{-6}$ radians is actually a measure of the in-flow speed at the sun's surface, and that gives $v=615$km/s, in agreement with   the numerical value computed for $v$ at the surface of the sun from (\ref{eqn:E4}).

These generalised Maxwell equations also predict gravitational lensing produced by the large in-flows, in (\ref{eqn:vexactb}), that are the  new `black holes' in galaxies.  Until now these anomalously large lensings have been also attributed,  using GR, to the presence of `dark matter'.  One example is reported in \cite{DMGalaxies} and another in \cite{JeeDMRing} which is re-analaysed without requiring dark matter in \cite {DMRing}.

\section{Generalised  Schr\"{o}dinger Equation: Emergent Gravity and Equivalence Principle \label{sect:acceln}}

A  generalisation of the  Schr\"{o}dinger equation is also required   \cite{Schrod}:
\begin{equation}
i\hbar\frac{\partial  \psi({\bf r},t)}{\partial t}=H(t)\psi({\bf r},t),
\label{eqn:equiv7}\end{equation}
where the free-fall hamiltonian is uniquely
\begin{equation}
H(t)=-i\hbar\left({\bf
v}.\nabla+\frac{1}{2}\nabla.{\bf v}\right)-\frac{\hbar^2}{2m}\nabla^2
\label{eqn:equiv8}\end{equation}
This follows from  the wave function being attached to the dynamical 3-space, and not to the embedding space, and that $H(t)$ be hermitian. We can compute the acceleration of a localised wave packet  using the Ehrenfest method \cite{Schrod}, and we obtain
\begin{equation}{\bf g}\equiv\frac{d^2}{dt^2}\left(\psi(t),{\bf r}\psi(t)\right)  
=\frac{\partial{\bf v}}{\partial t}+({\bf v}.\nabla){\bf v}+
(\nabla\times{\bf v})\times{\bf v}_R+...
\label{eqn:E11}\end{equation}
where ${\bf v}_R={\bf v}_0-{\bf v}$  is the velocity of the wave packet relative to the local space, as ${\bf v}_0$ is  the velocity relative to the embedding space. The vorticity term  causes rotation of the wave packet. For this to occur (\ref{eqn:E1}) must be generaalised to the case of non-zero vorticity \cite{Book}. This vorticity effect explains the Lense-Thirring effect, and such vorticity  is being detected by the Gravity Probe B satellite gyroscope experiment \cite{GPB}. We see, as promised, that this quantum-matter acceleration is equal to that of the 3-space itself, as in (\ref{eqn:E3}). This is the first derivation of the phenomenon of gravity from a deeper theory: gravity is a quantum effect - namely the refraction of quantum waves by the internal differential motion of the substructure  patterns to 3-space itself. Note that the equivalence principle has now been explained, as this `gravitational' acceleration is independent of the mass $m$ of the quantum system. 

\section{Generalised  Dirac Equation:  Relativistic Effects in 3-Space}
An analogous generalisation of the Dirac equation is also necessary giving the coupling of the spinor to the actual dynamical 
3-space, and again not to the embedding space as has been the case up until now: 
\begin{equation}
i\hbar\frac{\partial \psi}{\partial t}=-i\hbar\left(  c{\vec{ \alpha.}}\nabla + {\bf
v}.\nabla+\frac{1}{2}\nabla.{\bf v}  \right)\psi+\beta m c^2\psi
\label{eqn:12}\end{equation}
where $\vec{\alpha}$ and $\beta$ are the usual Dirac matrices. Repeating the analysis in (\ref{eqn:E11}) for the 3-space-induced acceleration we obtain
\begin{equation}\label{eqn:E12}
{\bf g}=\displaystyle{\frac{\partial {\bf v}}{\partial t}}+({\bf v}.{\bf \nabla}){\bf
v}+({\bf \nabla}\times{\bf v})\times{\bf v}_R-\frac{{\bf
v}_R}{1-\displaystyle{\frac{{\bf v}_R^2}{c^2}}}
\frac{1}{2}\frac{d}{dt}\left(\frac{{\bf v}_R^2}{c^2}\right)+...
\label{eqn:E13a}\end{equation}
which generalises  (\ref{eqn:E11}) by having a term which limits the speed of the wave packet relative to 3-space, $|{\bf v}_R|$, to be $<\!c$. This equation specifies the trajectory of a spinor wave packet in the dynamical 3-space.  The last term causes elliptical orbits
 to precess - for circular orbits $|{\bf v}_R|$ is independent of time.

\section{Deriving the Spacetime Geodesic Formalism: Local Poincar\'{e} Symmetry\label{section:spacetime}}

 We find that (\ref{eqn:E12}) may be also obtained by extremising the time-dilated elapsed time 
\begin{equation}
\tau[{\bf r}_0]=\int dt \left(1-\frac{{\bf v}_R^2}{c^2}\right)^{1/2}
\label{eqn:E13}\end{equation}  
with respect to the wave-packet trajectory ${\bf r}_0(t)$ \cite{Book}. This happens because of the Fermat least-time effect for waves: only along the minimal time trajectory do the quantum waves  remain in phase under small variations of the path. This again emphasises  that gravity is a quantum matter wave  effect.   We now introduce an effective  spacetime mathematical construct according to the metric
\begin{equation}
ds^2=dt^2 -(d{\bf r}-{\bf v}({\bf r},t)dt)^2/c^2 
=g_{\mu\nu}dx^{\mu}dx^\nu
\label{eqn:E14}\end{equation}
which is of the Panlev\'{e}-Gullstrand class of metrics \cite{Panleve,Gullstrand}. Then we have a  Local Poinacr\'{e} Symmetry, namely the transformations that leave  $ds^2$ locally invariant under a change of coordinates.   As well wave effects from (\ref{eqn:E1}) cause `ripples' in this induced spacetime, giving a different account of gravitational waves.  
The elapsed time in (\ref{eqn:E13}) may then be written as 
\begin{equation}
\tau=\int dt\sqrt{g_{\mu\nu}\frac{dx^{\mu}}{dt}\frac{dx^{\nu}}{dt}}.
\label{eqn:E14b}\end{equation}
The minimisation of  (\ref{eqn:E14b}) leads to the geodesics of the spacetime, which are thus equivalent to the trajectories from (\ref{eqn:E13}), namely (\ref{eqn:E13a}). 
We may introduce the  standard differential geometry curvature tensor for the induced  spacetime
\begin{equation}
R^\rho_{\mu\sigma\nu}=\Gamma^\rho_{\mu\nu,\sigma}-\Gamma^\rho_{\mu\sigma,\nu}+
\Gamma^\rho_{\alpha\sigma}\Gamma^\alpha_{\mu\nu}-\Gamma^\rho_{\alpha\nu}\Gamma^\alpha_{\mu\sigma},
\label{eqn:curvature}\end{equation}
where $\Gamma^\alpha_{\mu\sigma}$ is the affine connection for the metric in (\ref{eqn:E14})
\begin{equation}
\Gamma^\alpha_{\mu\sigma}=\frac{1}{2} g^{\alpha\nu}\left(\frac{\partial g_{\nu\mu}}{\partial x^\sigma}+
\frac{\partial g_{\nu\sigma}}{\partial x^\mu}-\frac{\partial g_{\mu\sigma}}{\partial x^\nu} \right).
\label{eqn:affine}\end{equation}
with  $g^{\mu\nu}$  the matrix inverse of  $g_{\mu\nu}$. 
In this formalism the trajectories of quantum-matter wave-packet test objects are determined by
\begin{equation}
\frac{d^2x^\lambda}{d\tau^2}+\Gamma^\lambda_{\mu\nu}\frac{dx^\mu}{d\tau}\frac{dx^\nu}{d\tau}=0,
\label{eqn:33}\end{equation}
as this is equivalent to (\ref{eqn:E12}).  In the standard treatment of GR  the geodesic for classical matter in  (\ref{eqn:33}) is a definition, and has no explanation.  Here we see that it is finally derived, but as a quantum matter effect.
Hence by coupling the Dirac spinor dynamics to the dynamical 3-space  we derive the geodesic formalism of General Relativity as a quantum effect, but without reference to the Hilbert-Einstein equations for the induced metric.  Indeed in general the metric of  this induced spacetime will not satisfy  these equations as the dynamical space involves the $\alpha$-dependent  dynamics, and $\alpha$ is missing from GR.  
We can also define the Ricci curvature scalar 
\begin{equation} R=g^{\mu\nu}R_{\mu\nu}\label{eqn:Ricci}\end{equation} 
where $R_{\mu\nu}=R^\alpha_{\mu\alpha\nu}$.  In general the induced spacetime in (\ref{eqn:E14}) has a non-zero Ricci scalar - it is a curved spacetime. We shall compute the Ricci  scalar for the expanding 3-space solution below.

We can also derive the Schwarzschild metric without reference to GR.  To do this we merely have to identify the induced spacetime metric corresponding to the in-flow in (\ref{eqn:E4}) outside of a spherical matter system, such as the earth.  Then (\ref{eqn:E14})  becomes
 \begin{equation}
ds^2=dt^{ 2}-\frac{1}{c^2}(dr+\sqrt{\frac{2GM(1+\frac{\alpha}{2}+..)}{r}}dt)^2
-\frac{r^2}{c^2}(d\theta^{ 2}+\sin^2(\theta)d\phi^2)
\label{eqn:GRE15}\end{equation}
 Making the change of variables $t\rightarrow t^\prime$ and
$\bf{r}\rightarrow {\bf r}^\prime= {\bf r}$ with
\begin{eqnarray}
t^\prime=&& t-
\frac{2}{c}\sqrt{\frac{2 GM(1{+}\frac{\alpha}{2}{+}\dots)r}{c^2}}+  \nonumber \\
&&\frac{4\ GM(1{+}\frac{\alpha}{2}{+}\dots)}{c^3}\,\mbox{tanh}^{-1}\sqrt{\frac{2 GM(1{+}\frac{\alpha}{2}{+}\dots)}{c^2r}}
\label{eqn:GRE16}\end{eqnarray}
this becomes (and now dropping the prime notation)
\begin{eqnarray}
ds^2&=&\left(1-\frac{2GM(1+\frac{\alpha}{2}+..)}{c^2r}\right)dt^{ 2} 
-\frac{1}{c^2}r^{ 2}(d\theta^2+\sin^2(\theta)d\phi^2) \nonumber \\
&&-\frac{dr^{ 2}}{c^2\left(1-{\displaystyle\frac{
2GM(1+\frac{\alpha}{2}+..)}{ c^2r}}\right)}.
\label{eqn:GRE17}\end{eqnarray}
which is  one form of the the Schwarzschild metric but with the $\alpha$-dynamics induced effective mass shift. Of course this is only valid outside of the spherical matter distribution, as that is the proviso also on (\ref{eqn:E4}). 
Hence in the case of the Schwarzschild metric the dynamics missing from both the Newtonian theory of gravity and General Relativity is merely hidden in a mass redefinition, and so didn't affect the various standard tests of GR, or even of Newtonian gravity.   A non-spherical symmetry version of the Schwarzchild metric is used in modelling the Global Positioning System (GPS).

\section{Supernova and Gamma-Ray-Burst Data} 

In the next section we show that the 3-space dynamics in (\ref {eqn:E1}) has an expanding space solution.
The supernovae and gamma-ray bursts provide standard candles that enable observation  of the expansion of the universe.  To test yet further that dynamics we compare the predicted expansion against the observables, namely the magnitude-redshift data from supernovae and gamma-ray bursts.
 The supernova data set used herein and shown in Figs. \ref{fig:SN1} and \ref{fig:SN2} is available at \cite{data set}.    Quoting from  \cite{data set}  we note that Davis {\it et al.} \cite{Davis}  combined several data sets by taking  the ESSENCE data set from Table 9 of Wood--Vassey {\it et al.}  (2007) \cite{WV}, using only the supernova that passed the light-curve-fit quality criteria. They took the HST data from Table 6 of Riess {\it et al.} (2007) \cite{Riess}, using only the supernovae classified as gold.
To put these data sets on the same Hubble diagram  Davis {\it et al.} used 36 local supernovae that are in common between these two data sets. When discarding supernovae with $z<0.0233$ (due to larger uncertainties in the peculiar velocities) they found an offset of $0.037 \pm 0.021$ magnitude between the data sets, which they then corrected for by subtracting this constant from the HST data set. The dispersion in this offset was also accounted for in the uncertainties.
The HST data set had an additional 0.08 magnitude added to the distance modulus errors to allow for the intrinsic dispersion of the supernova luminosities. The value used by Wood--Vassey {\it et al.}  (2007) \cite{WV} was instead 0.10 mag. Davis {\it  et al.}  adjusted for this difference by putting the Gold supernovae on the same scale as the ESSENCE supernovae. Finally, they also added the dispersion of 0.021 magnitude introduced by the simple offset described above to the errors of the 30 supernovae in the HST data set. The final supernova data base for  the distance modulus $\mu_{obs}(z)$ is shown in Figs. \ref{fig:SN1} and \ref{fig:SN2}.  The gamma-ray-burst (GRB) data is from Schaefer \cite{GRB}.

\section{Expanding Universe from Dynamical 3-Space}

Let us now explore the expanding  3-space  from (\ref {eqn:E1}).  Critically, and unlike the GR-FLRW model, the 3-space expands even when the energy density is zero.
Suppose that  we have a radially symmetric effective density $\rho(r,t)$, modelling EM radiation, matter, cosmological constant etc,   and that we look for a radially symmetric time-dependent flow ${\bf v}({\bf r},t) =v(r,t)\hat{\bf r}$ from (\ref{eqn:E1}) (with $\beta=-\alpha$).  Then $v(r,t)$ satisfies the equation,  with $v^\prime=\displaystyle{\frac{\partial v(r,t)}{\partial r}}$,
\begin{equation}
\frac{\partial}{\partial t}\left( \displaystyle{\frac{2v}{r}}+v^\prime\right)+vv^{\prime\prime}+2\frac{vv^{\prime}}{r}+ (v^\prime)^2+\frac{\alpha}{4}\left(\frac{v^2}{r^2} +\frac{2vv^\prime}{r}\right)
=- 4\pi G \rho(r,t)  \label{eqn:radialflow}\end{equation}
Consider first the zero energy case $\rho=0$. Then we have a Hubble  solution $v(r,t)=H(t)r$, a centreless flow, determined by
\begin{equation}{\dot H}+\left(1+\frac{\alpha}{4}\right)H^2=0
\end{equation}
with ${\dot H}=\displaystyle{\frac{dH}{dt}}$.  We also introduce in the usual manner the scale factor $a(t)$ according to $H(t)=\displaystyle{\frac{1}{a}\frac{da}{dt}}$. We then obtain
the solution
\begin{equation}
H(t)=\frac{1}{(1+\frac{\alpha}{4})t}=H_0\frac{t_0}{t}; \mbox{\ \  }  a(t)=a_0\left(\frac{t}{t_0} \right)^{4/(4+\alpha)}
\label{eqn:spacexp}\end{equation}
where $H_0=H(t_0)$ and $a_0=a(t_0)$.  Note that we obtain an expanding 3-space even where the energy density is zero - this is in sharp contrast to the GR-FLRW model for the expanding universe, as shown below.

We can write the  Hubble function $H(t)$ in terms of $a(t)$ via the inverse function $t(a)$, i.e. $H(t(a))$ and finally as $H(z)$, where the redshift observed now, $t_0$, relative to the wavelengths at time $t$, is  $z=a_0/a-1$. Then we obtain
\begin{eqnarray}
H(z)={H_0}(1+z)^{1+\alpha/4}
\label{eqn:H2a}\end{eqnarray}
To test this expansion we need to predict the relationship between the cosmological observables, namely the relationship between the apparent energy-flux magnitudes and redshifts. This  involves taking account of the reduction in photon count caused by the expanding 3-space, as well as the accompanying reduction in photon energy. To that end we first
 determine the distance travelled by the light from a supernova or GRB  event before detection.  Using a choice of embedding-space coordinate system with $r=0$ at the location of a supernova/GRB event  the 
speed of light relative to this embedding space frame is $c+v(r(t),t)$, i.e. $c$ wrt the space itself, as noted above,  where $r(t)$ is the embedding-space distance from the source. Then the distance travelled by the light at time $t$ after emission at time $t_1$ is determined implicitly by
\begin{equation}
r(t)=\int_{t_1}^t dt^\prime(c+v(r(t^\prime), t^\prime),
\label{eqn:distance1}\end{equation}
which has the solution on using $v(r,t)=H(t)r$
\begin{equation}
r(t)=c a(t)\int_{t_1}^t \frac{dt^\prime}{a(t^\prime)}.
\label{eqn:distance2}\end{equation}
This distance gives directly the surface area $4\pi r(t)^2$ of  the expanding sphere  and so the decreasing photon count per unit of that surface area. However  also because of the expansion the flux of photons is reduced by the factor  $1/(1+z)$, simply because they are spaced further apart by the expansion. The photon flux is then given by
\begin{equation}
{\cal F}_P=\frac{{\cal L}_P}{4\pi r(t)^2(1+z)}
\end{equation}
where  ${\cal L}_P$ is the source photon-number luminosity. However usually the energy flux is measured, and the energy of each photon is reduced by the factor $1/(1+z)$ because of the redshift. Then the energy flux is, in terms of the source energy luminosity ${{\cal L}_E}$, 
\begin{equation}
{\cal F}_E=\frac{{\cal L}_E}{4\pi r(t)^2(1+z)^2}\equiv\frac{{\cal L}_E}{4\pi r_L(t)^2}
\end{equation}
which  defines the effective energy-flux luminosity distance $r_L(t)$.
Expressed in terms of the observable redshift $z$ this gives an energy-flux  luminosity effective distance
\begin{equation}
r_L(z)=(1+z)r(z)=c (1+z)\int_{0}^z \frac{dz^\prime}{H(z^\prime)}
\label{eqn:distance3}\end{equation}
The dimensionless `energy-flux'  luminosity effective distance is then given by
 \begin{equation}
d_L(z)=(1+z)\int_0^z \frac{H_0 dz^\prime}{H(z^\prime)}
\label{eqn:H1a}\end{equation}
 and the theory distance modulus is  defined by
\begin{equation}
\mu(z)=5\log_{10}(d_L(z))+m.
\label{eqn:H1b}\end{equation}
Because all the selected supernova have the same absolute magnitude, $m$ is a constant whose value is determined by fitting the low $z$ data. The GRB magnitudes have been adjusted to match the supernovae data \cite{GRB}.

Using the  Hubble expansion (\ref{eqn:H2a}) in (\ref{eqn:H1a}) and (\ref{eqn:H1b}) we obtain the middle curves (red) in Figs. \ref{fig:SN1} and  the \ref{fig:SN2}, yielding an excellent agreement with the supernovae and GRB data. Note that because $\alpha/4$ is so small it actually has negligible effect on these plots.  But that is only the case for the homogeneous expansion - we saw above that the $\alpha$ dynamics can result in large effects such as black holes and large spiral galaxy rotation effects when the 3-space is inhomogeneous.  Hence the dynamical 3-space gives an immediate account of the universe expansion data, and does not require the introduction of  a cosmological constant or `dark energy', but which will be nevertheless discussed next. 

\begin{figure}
\vspace{0mm}
\hspace{15mm}\includegraphics[scale=0.6]{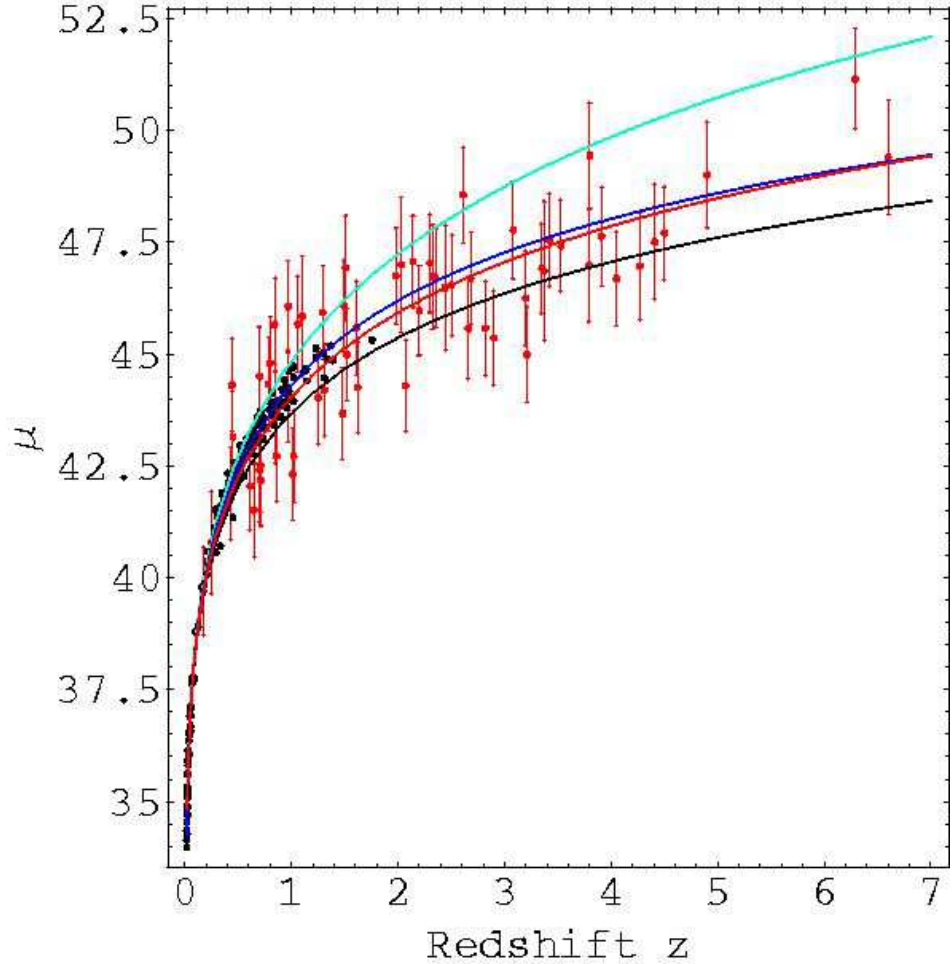}
\vspace{-5mm}\caption{\small{ Hubble diagram showing the combined supernovae data from Davis {\it et al.} \cite{Davis} using several data sets from   Riess {\it et al.} (2007)\cite{Riess} and Wood-Vassey {\it et al.}  (2007)\cite{WV} (dots without error bars for clarity - see figure \ref{fig:SN2} for error bars) and the Gamma-Ray-Bursts data (with error bars) from Schaefer \cite{GRB}.  Upper curve (green)  is `dark energy' only $\Omega_\Lambda=1$, lower curve  (black) is matter only $\Omega_m=1$. Two middle curves show best-fit of `dark energy'-`dark-matter' (blue) and dynamical 3-space prediction (red), and are essentially indistinguishable.  However the theories make very different predictions for the future. We see that the best-fit `dark energy'-`dark-matter' curve essentially converges on the uniformly-expanding parameter-free dynamical 3-space prediction. See figure \ref{fig:difference} for comparison out to $z=14$.}
\label{fig:SN1}}\end{figure}

\begin{figure}
\vspace{3.0mm}
\hspace{15mm}{\includegraphics[scale=0.6]{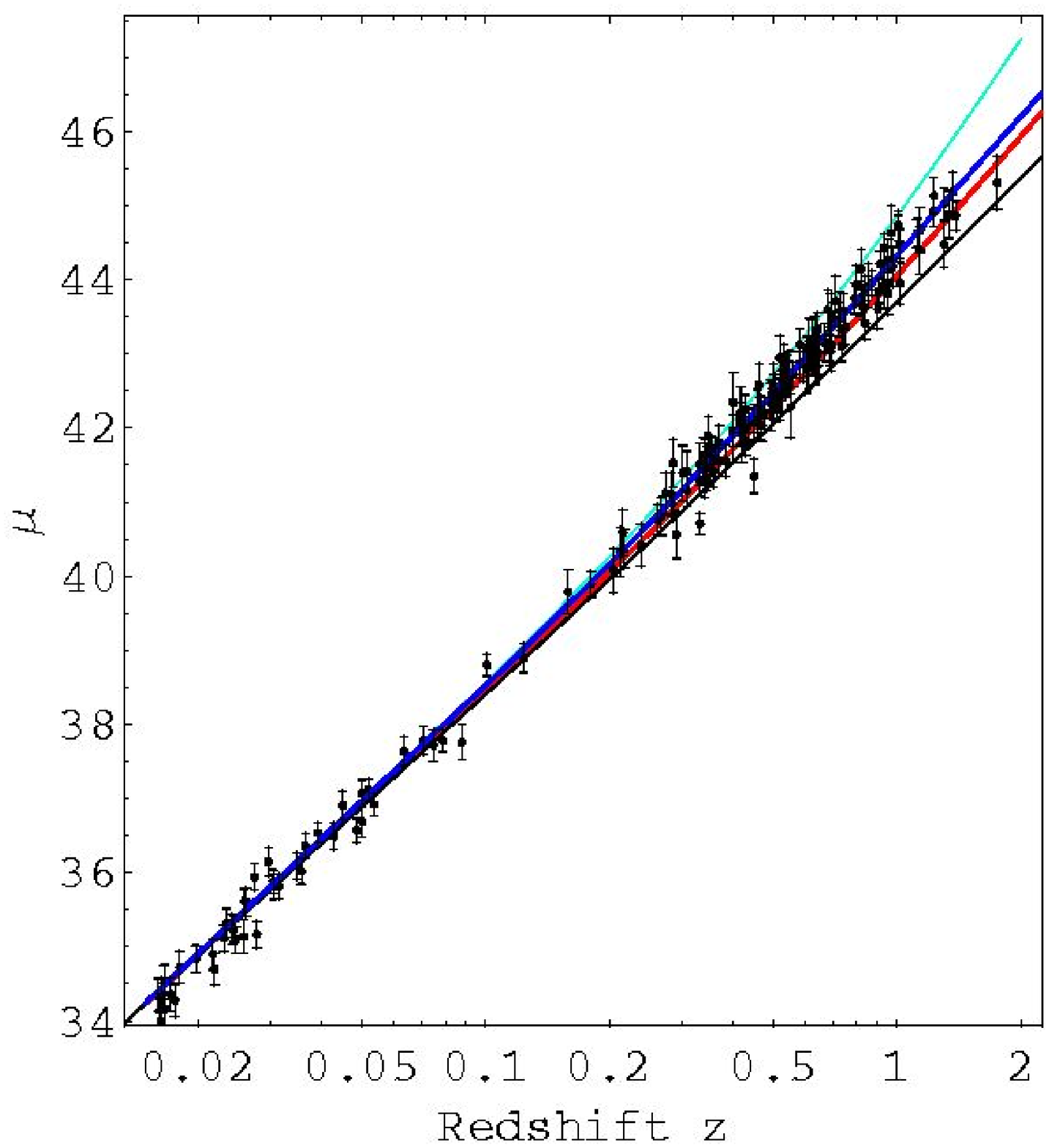}}\,
{\caption{\small{  Hubble diagram as in figure \ref{fig:SN1} but plotted logarithmically  to reveal details for  $z<2$, and without GRB data.  Upper curve (green) is `dark energy' only $\Omega_\Lambda=1$. Next curve (blue) is  best fit of `dark energy'-`dark-matter'. Lowest curve (black) is matter only $\Omega_m=1$. 2nd lowest curve (red) is  dynamical 3-space prediction. }\label{fig:SN2}}}
\end{figure}

\section{Expanding Universe - Non-Zero Energy Density Case}
When the energy density is not zero we need to take account of the dependence of $\rho(r,t)$ on the scale factor of the universe. In the usual manner we thus write
\begin{equation}
\rho(r,t)=\frac{\rho_{m}}{a(t)^3}+\frac{\rho_{r}}{a(t)^4}+\Lambda
\label{eqn:evolve}\end{equation}
for matter, EM radiation and the cosmological constant or `dark energy' $\Lambda$, respectively, where the matter and radiation is approximated by a spatially uniform (i.e independent of $r$)  equivalent matter density. We argue here that $\Lambda$ - the cosmological constant or dark energy density, like dark matter, is an unnecessary concept. We have chosen a definition for the cosmological constant $\Lambda$ so that it has the units of matter density.
Then (\ref{eqn:radialflow}) becomes for $a(t)$
\begin{eqnarray}
\frac{\ddot  a}{a}+\frac{\alpha}{4}\frac{{\dot a}^2}{a^2}
=-\frac{4\pi G}{3}\left(\frac{\rho_{m}}{a^3}+\frac{\rho_{r}}{a^4}+\Lambda \right)
\label{eqn:Reqn}\end{eqnarray}giving
\begin{equation}
{\dot a}^2=\frac{8\pi G}{3}\left(\frac{\rho_{m}}{a}+\frac{\rho_{r}}{a^2}+\Lambda a^2\right)-\frac{\alpha}{2}\int\frac{{\dot a}^2}{a}da+f
\label{eqn:R2}\end{equation}
where $f$ is an integration constant.
 In terms of ${\dot a}^2$ this has the solution
\begin{equation}{\dot a}^2\!=\!\frac{8\pi G}{3}\!\!\left(\!\frac{\rho_{m}}{(1-\frac{\alpha}{2})a}\!+\!\frac{\rho_{r}}{(1-\frac{\alpha}{4})a^2}\!+\!\frac{\Lambda a^2}{(1+\frac{\alpha}{4})}\!+\!b a^{-\alpha/2}\!\right)
\label{eqn:R3}\end{equation}
which is easily checked by substitution into (\ref{eqn:R2}), and where $b$ is the  integration constant. Finally we obtain from  (\ref{eqn:R3})
\begin{equation}
t(a)=t(a_0)+\int^a_{a_0}\frac{da}{\sqrt{\displaystyle{\frac{8\pi G}{3}}\left(\displaystyle{\frac{\rho_{m}}{a}+\frac{\rho_{r}}{a^2}}+\Lambda a^2+b a^{-\alpha/2}\right)}}
\label{eqn:R4}\end{equation} 
where  we have re-scaled the various density parameters for notational convenience.  When $\rho_m=\rho_r=\Lambda=0$, (\ref{eqn:R4})
reproduces the expansion in (\ref{eqn:spacexp}), and so the density terms in (\ref{eqn:R3}) give the modifications to  the dominant purely spatial expansion, which we have noted above  already gives an excellent account of the data. It is important to note that  (\ref{eqn:R3}) has the  $b$ term  - the constant of integration, even when $\alpha=0$, whereas the GR-FLRW dynamics demands, effectively, $b=0$. Having $b\neq 0$ simply asserts that the 3-space can expand even when the energy density is zero - an effect missing from GR-FLRW cosmology.

From (\ref{eqn:R3})  we then obtain
\begin{equation}
H(z)^2={H_0}^2(\Omega_m(1+z)^3+\Omega_r(1+z)^4 +\Omega_\Lambda+\Omega_s(1+z)^{2+\alpha/2})
\label{eqn:H2}\end{equation}
where 
\begin{equation}
H_0=\left(\frac{8\pi G}{3}(\rho_m+\rho_r+\Lambda+b)\right)^{1/2}
\label{eqn:Hubble constant}\end{equation}
\begin{equation}
\Omega_m=\rho_m/(\rho_m+\rho_r+\Lambda+b),...
 \end{equation}
and so
\begin{equation}
\Omega_m+\Omega_r+\Omega_\Lambda+\Omega_s=1.
\end{equation}

Next we discuss  the strange feature of the  GR-FLRW dynamics which requires a non-zero energy density for the universe to expand.

\begin{figure}
\hspace{-5mm}{\includegraphics[scale=0.5]{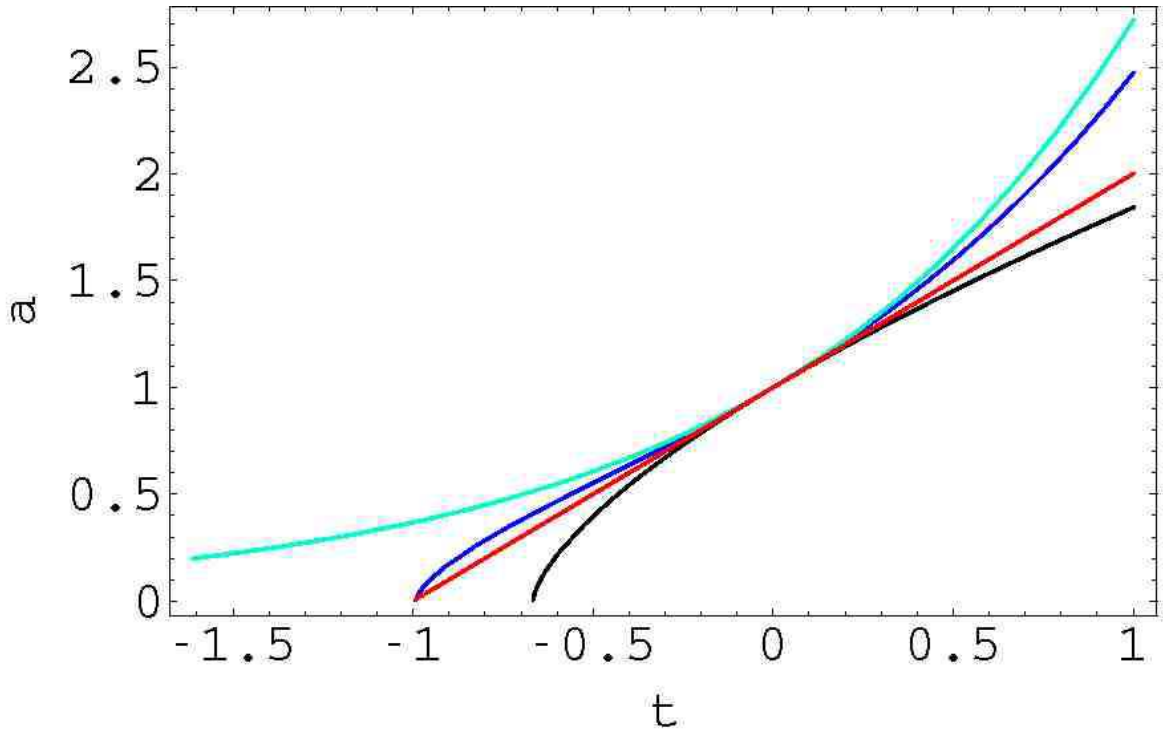}}\,
{\vspace{-4mm}\caption{\small{  Plot of the scale factor $a(t)$ vs $t$, with $t=0$ being `now' with $a(0)=1$, for the four cases discussed in the text, and corresponding to the plots in  Figs. \ref{fig:SN1} and \ref{fig:SN2}: (i) the upper  curve (green) is the `dark energy' only case, resulting in an exponential acceleration at all times, (ii) the bottom curve (black) is the matter only prediction, (iii) the 2nd highest curve (to the right of $t=0$) is the fitted  `dark energy' plus `dark-matter 'case (blue) showing a past deceleration and future exponential acceleration effect. The straight line plot  (red)  is the dynamical 3-space prediction.  We see that the best-fit `dark energy' - `dark matter' curve closely follows  the dynamical 3-space result. All plots have the same slope at $t=0$, i.e. the same value of $H_0$. 
 }\label{fig:Rtplot}}}
\end{figure}

\section{Deriving the Friedmann-Lema\^{i}tre-Robertson-Walker Metric}
The induced effective spacetime metric in (\ref{eqn:E14}) is, for the Hubble expansion,
\begin{equation}
ds^2=g_{\mu\nu}dx^\mu dx^\nu=dt^2-(d{\bf r}-H(t){\bf r}dt)^2/c^2 
\label{eqn:PGmetric}\end{equation}
 The occurrence of $c$ has nothing to do with the dynamics of the 3-space - it is related to the geodesics of relativistic quantum matter, as noted above. Nevertheless changing to spatial coordinate variables   ${\bf r}^\prime$ with  ${\bf r}=a(t){\bf r}^\prime$, and with $t^\prime=t$,  we obtain
\begin{equation}
ds^2=g_{\mu\nu}dx^\mu dx^\nu=dt^{\prime2}-a(t^\prime)^2d{\bf r}^{\prime2}/c^2 
\label{eqn:Hubblemetric}\end{equation}
which is the usual   Friedmann-Lema\^{i}tre-Robertson-Walker (FLRW) metric in the case of a flat spatial section.  However this involves a deceptive choice of spacetime coordinates.  Consider the position of a galaxy located at 
${\bf r}(t)$. Then over the time interval $dt$ this galaxy moves a distance $dr={\bf v}({\bf r},t)dt=H(t){\bf r}(t)dt$. In terms of the FLRW distance however the galaxy moves through distance $d{\bf r}^\prime=d({\bf r}(t)/a(t))=(d{\bf r}(t)-H(t){\bf r}(t))/a(t)={\bf 0}$. Hence the FLRW distances involve a dynamically determined re-scaling of the spatial distance measure so that the universe does not expand in terms of these coordinates.   We now show why the GR-FLRW cosmology model needs to invoke `dark energy' and `dark matter' to fit the observational data.The Hilbert-Einstein (HE) equations for a spacetime metric are 
\begin{equation}
G_{\mu\nu}\equiv R_{\mu\nu}-\frac{1}{2}Rg_{\mu\nu}=8\pi  G\Lambda g_{\mu\nu}+8\pi G T_{\mu\nu}
\label{eqn:HE}\end{equation}
where $G_{\mu\nu}$ is supposed to describe the dynamics of the spacetime manifold in the presence of an energy-momentum described by the  tensor $T_{\mu\nu}$. Surprisingly, in the absence of $\Lambda$ and $T_{\mu\nu}$  the HE equation, now $G_{\mu\nu}=0$, does not have an expanding universe solution for the metric in (\ref{eqn:Hubblemetric}).

The stress-energy tensor is, according to the Weyl postulate,
 \begin{equation}
T_{\mu\nu}=(\rho+p)u_\mu u_\nu+pg_{\mu\nu}
\label{eqn:stress}\end{equation}
Then with $u^\mu=(1,0,0,0)$ we obtain  for the flat spacetime in (\ref{eqn:Hubblemetric}) the well-known Friedmann equations
\begin{equation}
\frac{{\dot a}^2}{a^2}=\frac{8\pi G\Lambda}{3}+\frac{8}{3}\pi G \rho
\label{eqn:HE1}\end{equation}
\begin{equation}
\frac{{\ddot a}}{a}+\frac{{\dot a}^2}{2a^2}=4\pi G\Lambda-4\pi G p
\label{eqn:HE2}\end{equation}
These two equations constitute the dynamical equations for the current standard model of cosmology ($\Lambda$CDM).  Even in the case of zero-pressure `dust', with $p=0$, these two equations are not equivalent to (\ref{eqn:Reqn}) (with $\alpha=0$ in this section).  If $\rho=0, \Lambda=0$ and $p=0$ then these equations give the non-expanding universe $\dot a=0$, which is not the general solution to  (\ref{eqn:Reqn}) which has $\dot a$= constant, and   it is this solution which gives a parameter-free fit to the supernova/GRB redshift data. If only $p=0$ then these two equations give, first from (\ref{eqn:HE1}), and then  from  (\ref{eqn:HE1}) and  (\ref{eqn:HE2}).
\begin{equation}
\frac{{\dot a}^2}{2}-\frac{4\pi G\Lambda a^2}{3}-\frac{4\pi G \rho_m}{3a}=0
\label{eqn:p1}\end{equation}
\begin{equation}
\frac{d}{dt}\left(\frac{{\dot a}^2}{2}-\frac{4\pi G\Lambda a^2}{3}-\frac{4\pi G \rho_m}{3a}\right)=0
\label{eqn:p2}\end{equation}
Whence (\ref{eqn:p1}) requires that the integration constant from  (\ref{eqn:p2}) must be zero - this is equivalent to demanding $b=0$ in (\ref{eqn:R3}).  Hence according to the GR-FLRW dynamics the universe can only expand if at least one of $\Lambda$ or $\rho_m$ is non-zero.   This amounts to not modelling space itself as a dynamical system - only the relative motion of energy/matter has any ontological meaning: this has been the main theme of spacetime modeling from the beginning.  In dealing with this failure of the GR-FLRW dynamics we now show that a judicious choice of  $\Omega_\Lambda$ and  $\Omega_m$ can mock up the 3-space expansion, but only by introducing an extraneous and spurious acceleration.

\section{Predicting the $\Lambda$CDM  Parameters  $\Omega_\Lambda$ and $\Omega_{DM}$}

\begin{figure}
\vspace{0mm}
\hspace{-2mm}\includegraphics[scale=0.5]{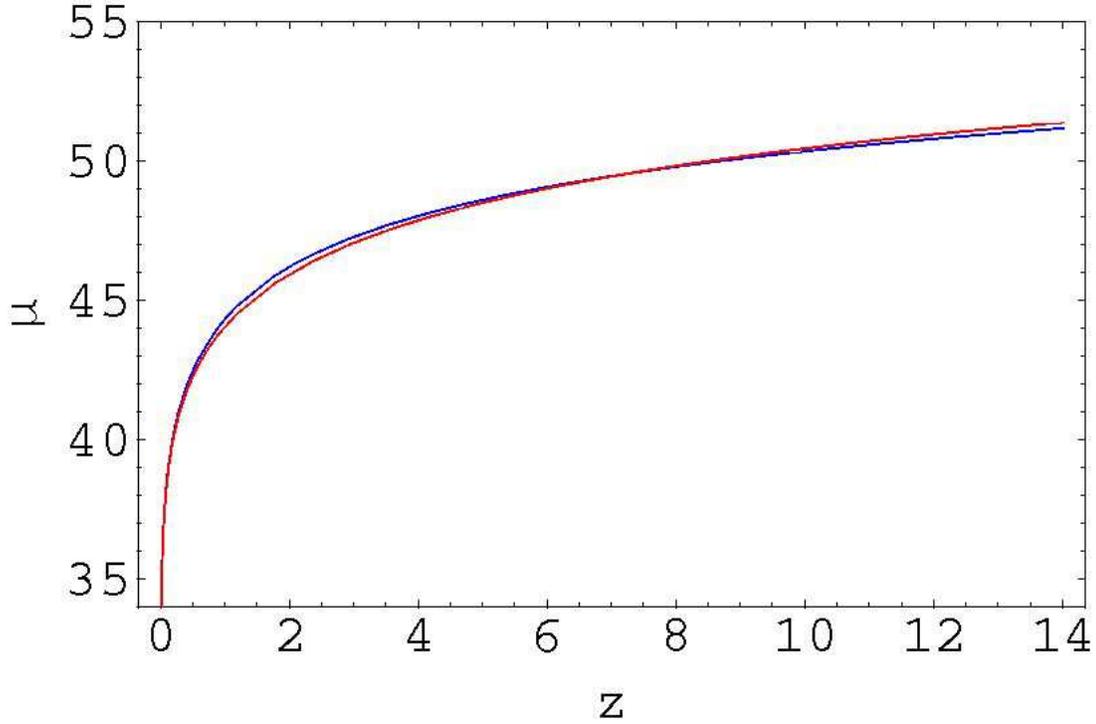}
\vspace{-1mm}\caption{\small{ Comparison of GR-FLRW distance modulus $\mu(z)$ with $\Omega_\Lambda=0.73, \Omega_m=1-\Omega_\Lambda=0.27$, (blue plot), with distance modulus from dynamical 3-space (red plot). The small difference, which could not be distinguished by the observational data, over this redshift range demonstrates that the GR-FLRW model simulates the parameter-free uniformly-expanding dynamical 3-space prediction.  Hence  the `standard model'  values $\Omega_\Lambda=0.73$, $\Omega_m=0.27$ are predictable without reference to the actual supernovae/GRB magnitude-redshift data - there is no need to invoke `dark energy' nor `dark matter'. The GR-FLRW model does not permit an expanding space unless there is energy residing in the space.}
\label{fig:difference}}\end{figure}

It is argued herein that `dark energy' and `dark matter' arise in the GR-FLRW cosmology because in that model space cannot expand unless there is an energy density present in the space, if that space is flat and the energy density is pressure-less. Then essentially fitting the Friedmann  model  $\mu(z)$ to the dynamical 3-space cosmology $\mu(z)$ we obtain 
 $\Omega_\Lambda=0.73$, and so $\Omega_m=1-\Omega_\Lambda=0.27$. These values arise from a best fit for $z\in\{0,14\}$, and the quality of the fit is shown in figure \ref{fig:difference}.  The actual values for $\Omega_\Lambda$ depend on the red-shift range used, as the Hubble functions for the GR-FLRW and dynamical 3-space have different functional dependence on $z$. These values are of course independent of the actual observed redshift data. In fitting the Friedmann dynamics to the supernovae/GRB magnitude-redshift data the  best fit is  $\Omega_\Lambda=0.73$, and so $\Omega_m=0.27$ \cite{DETask}, p40. Of course since this amount of matter is much larger than the observed baryonic matter, it is claimed that most of this matter is  the so-called `dark matter'. Essentially the current standard model of cosmology $\Lambda$CDM  is excluded from modelling a uniformly expanding dynamical 3-space, but by choice of the parameter $\Omega_\Lambda$ the Hubble function $H_F(z)$ can be made to fit the data. However $H_F(z)$ has the wrong functional form; when applied to the future expansion of the universe the Friedmann dynamics  produces a spurious  exponentially expanding universe.

\section{Implications of the Supernovae and Gamma-Ray-Burst Data}

As already noted above the supernovae and gamma-ray-burst data show that the universe is uniformly expanding, and that such an expansion  cannot be produced by the Friedmann GR dynamics  for a flat 3-space except by a judicious choice of the parameters $\Omega_\Lambda$ and $\Omega_m=1-\Omega_\Lambda$.   Nevertheless we find that the FLRW flat 3-space spacetime metric is relevant but that it does not satisfy the Friedmann equations. We shall now illustrate this by comparing the distance moduli from various choices of the density parameters in (\ref{eqn:H2}).  We consider four choices of parameter values  with  the plots shown in  Figs. \ref{fig:SN1} and \ref{fig:SN2}:

 (i) A pure `dark energy' or cosmological constant driven expansion has $\Omega_m=0,  \Omega_r=0, \Omega_\Lambda=1, \Omega_s=0$.  This produces  a Hubble plot that causes too rapid an expansion, and indeed an exponential expansion at all epochs. This choice fails to fit the data.
 
 (ii)  A matter only expansion  has
$\Omega_m=1,  \Omega_r=0, \Omega_\Lambda=0, \Omega_s=0$. This produces a Hubble expansion that is de-accelerating and fails to fit the data.

 (iii)  The $\Lambda$CDM  Friedmann-GR parameters are  $\Omega_m=0.27,  \Omega_r=0, \Omega_\Lambda=0.73 , \Omega_s=0$. They arise from a fit to the dynamical 3-space uniformly-expanding  prediction as well as a best fit to the observational data. This shows that the data is implying a uniformly expanding 3-space. The Friedmann equations demand that $\Omega_s=0$ in the pressure-less dust case.
   
  (iv) The zero-energy dynamical 3-space has $\Omega_m=0,   \Omega_r=0,  \Omega_\Lambda=0,  \Omega_s=1$, as noted above. The spatial expansion dynamics alone gives a good account of the data. The data cannot distinguish between cases (iii) and (iv). 
  
  Of course the EM radiation term $\Omega_r$ is non-zero but small and determines the expansion during the baryongenesis initial phase, as does the spatial dynamics expansion term because of the $\alpha$ dependence.
  
  \section{Age of Universe and WMAP  Data}

The age of the universe is of course theory dependent.  From (\ref{eqn:R4}) it is given in general  by 
\begin{equation}
t_0=\int_0^1\frac{da}{\dot{a}(t)}=\int_0^\infty
\frac{dz}{(1+z)H(z)}
\label{eqn:Age}\end{equation} 
and so we must choose a form for $H(z)$, and one that models the redshift back to the Big Bang ($z=\infty$).  However we only have, at best, knowledge of $H(z)$ back to say $z \approx 7$. The GR-FLRW $H(z)$ essentially  fits to the 3-space form for $H(z)$ over a considerable range of $z$ values, as shown in figure \ref{fig:difference},  but not over the full $z$-range as shown in 
figure \ref{fig:Rtplot}. Indeed figure  \ref{fig:Rtplot} shows that the two $a(t)$ functions do differ, but that nevertheless they give essentially the same age for the universe. This is just an accident.  However as noted when applied to the future expansion  another extrapolation is employed and the GR-FLRW model predicts an exponential expansion, while the 3-space dynamics model predicts a continuing uniform expansion. From (\ref{eqn:H2a}), with $\alpha=0$, we obtain $t_0=1/H_0$.   However  there will be changes to this from including effects of baryonic matter and that when the universe is inhomogeneous $\rho_{DM}$ may not be small or even positive, and would not evolve as conserved matter does as in (\ref{eqn:evolve}).

 Analysis of the CMB anisotropies by WMAP \cite{WMAP1, WMAP2,Kosowsky}  have given results that are consistent with the $\Lambda$CDM model. However as noted herein that model involves a Hubble function that can also be matched by the Hubble function from the dynamical 3-space.  So the {\it concordance} between  fitting  the supernovae/GRB data and the CMB data to the $\Lambda$CDM model does not imply the correctness of   this model.  This issue has been discussed by Efstathiou and Brown \cite{Efstathiou}, and is known as the geometric degeneracy effect.  What is most telling in this context is more than the existence of this degeneracy effect, but that the $\Lambda$CDM model
 parameters can be accurately computed without reference to the observational data, so they are purely artifacts of using the GR-FLRW $\Lambda$CDM model.
 
   In this context we also note another geometric degeneracy, namely that if we use a FRW metric with a non-flat 3-space then the Friedmann equations now permit the term  with coefficient $b$ in  (\ref{eqn:R3}), but with $\alpha=0$, arises.   This term, however, has completely different origins: in the GR-FLRW cosmology it is associated with 3-space curvature, while above it is related to the dynamics of the {\it flat} 3-space.  
 
 So from the beginning of cosmology the flawed Friedmann model of an expanding universe with a non-dynamical 3-space has been employed.   The neglect of the 3-space dynamics up to now means that other methods for studying the so-called `dark energy' and `dark matter' need to be re-investigated: these include Baryonic Acoustic Oscillations (BAO),  Galaxy Cluster Counting (GCC) and Weak Gravitational Lensing  (WGL) \cite{DETask}.   In particular BAO analysis will be affected by the $\alpha$-dynamics term in (\ref{eqn:E1}) which can produce significant effects when the system is inhomogeneous. Similarly the GCC and WGL are also  affected by this $\alpha$-dynamics.  These effects impact on the determination of the baryonic matter content and on the computed age of the universe.
 
 \section{Ricci Curvature from the  Dynamical 3-Space}
 
 We now note the form of the Ricci scalar, which is a measure of the non-flatness of the induced spacetime metric. From either 
(\ref {eqn:PGmetric})  or (\ref{eqn:Hubblemetric}) we obtain the Ricci scalar to be
\begin{equation}
R=-6\left( \frac{{\dot{a}}^2}{a^2} +\frac{\ddot{a}}{a}\right) = \frac{-96+24\alpha}{(4+\alpha)^2t^2}\neq0
\label{eqn:RicciValue}\end{equation}
on using, say,  expression (\ref{eqn:spacexp}) for $a(t)$. So even though  the dynamical 3-space  leads to the FLRW  spacetime metric, with a flat 3-space, the spacetime itself is not flat.  Nevertheless it is important to note that the induced spacetime has no ontological significance - it is merely a mathematical construct.

\section{Conclusions}

We have argued that a self-limited  information-theoretic approach to modelling reality leads to an emergent quantum foam  formalism, and that this in turn can be modelled, in part,  by a classical dynamical 3-space. The minimal dynamics for this 3-space requires a two-parameter
dynamics for that 3-space, with one being $G$ and the other being $\alpha$. That this $\alpha$ is the fine structure constant is determined from various experimental/observational data.  Generalising the Schr\"{o}dinger and Dirac equations then explains the phenomenon of gravity - here gravity is an emergent phenomenon arising from the wave-nature of quantum matter, and is not simply modelled as a phenomenon with known properties.  The dynamical 3-space theory is then shown to explain various phenomena, including the so-called `dark matter' effects - essentially these are related to the  $\alpha$-dynamics that is missing from Newtonian gravity and GR.  The 3-space dynamics has an expanding flat-universe solution that gives a parameter-free account of the supernovae/GRB data.
This expansion occurs even when the energy density of the universe is zero. In contrast the GR-FLRW expansion dynamics only permits an expanding universe when the energy density, in the case of a pressure-less dust, is non-zero, and also essentially large.  To fit the expanding 3-space solution  a judicious  choice of  $\Omega_\Lambda=0.73$ and $\Omega_m=0.27$ in the GR-FLRW model is found, independent of the observational data.  Not surprisingly these are the exact values found from fitting the GR-FLRW dynamics to the supernovae/GRB data.  However a spurious aspect to this is that the GR-FLRW fit  generates an anomalous exponential expansion in the future, as the GR-FLRW Hubble function has the wrong functional form.
Because of the dominance of  $\Omega_\Lambda=0.73$ and $\Omega_m=0.27$ the GR-FLRW dynamics has become known as the $\Lambda$CDM `standard' model of cosmology.   It is thus argued that the Friedmann dynamics for the universe has been flawed from the very beginning of cosmology, and that the new high-precision supernova data has finally made that evident.   The derived theory of gravity does away with the need for  `dark energy' and `dark matter'.  As the dynamical 3-space theory is strongly non-local we see that the universe is more connected than in previous models, and that this offers a new explanation for the uniformity of the universe, i.e.   a resolution of  the so-called horizon problem.  The derivation of the phenomenon of gravity from  the deeper Quantum Homotopic Field Theory which, it is argued, arises from the information-theoretic model for reality.  The emergent gravity is seen to arise from a quantum system, and as well, displays features not in the Newtonian or the GR treatments of gravity. Hence the QHFT is a possible  quantum `theory of everything', including the formation of the universe via a phase transition.  It thus gives us a possible quantum cosmology, and which has a `physics' of the pre-universe, that is, before the beginning of a phase transition to a quasi-classical growing 3-space component.

\end{document}